\pdfoutput=1
\documentclass[a4paper]{article}
\usepackage{graphicx,epsfig,color}
\usepackage{jheppub}
\usepackage[english]{babel}
\usepackage{amssymb,amsfonts,amsmath}
\usepackage{verbatim}
\usepackage{physics}
\usepackage[dvipsnames]{xcolor}
\usepackage{bbold,bbm}
\newcommand{\be}{\begin{equation}} \newcommand{\ee}{\end{equation}}
\newcommand{\bea}{\begin{eqnarray}} \newcommand{\eea}{\end{eqnarray}}

\usepackage{cancel}

\usepackage{float}

\usepackage{booktabs}

\makeatletter
\def\@fpheader{\relax}
\makeatother

\newcommand{\mt}[1]{\textrm{\scriptsize #1}}
\def\Nc{N_\mt{c}}
\def\Nf{N_\mt{f}}
\def\ls{l_\mt{s}}
\def\gs{g_\mt{s}}
\def\gYM{g_\mt{YM}}

\begin{document}

\begin{flushright}HIP-2025-23/TH\end{flushright}
\title{Subtleties of non-Abelian D-brane actions and their effect on holographic heavy-light meson spectra}

\author[a,b]{Carlos Hoyos,}
\affiliation[a]{Department of Physics, Universidad de Oviedo, c/ Leopoldo Calvo Sotelo 18, ES-33007 Oviedo, Spain}
\affiliation[b]{Instituto de Ciencias y Tecnolog\'{\i}as Espaciales de Asturias, c/ Leopoldo Calvo Sotelo 18, ES-33007 Oviedo, Spain}
\emailAdd{hoyoscarlos@uniovi.es}

\author[c,d]{Niko Jokela,}
\affiliation[c]{Department of Physics, University of Helsinki, P.O. Box 64, FI-00014, University of Helsinki, Finland}
\affiliation[d]{Helsinki Institute of Physics, P.O. Box 64, FIN-00014 University of Helsinki, Finland}
\emailAdd{niko.jokela@helsinki.fi}

\author[e,f]{and Andrea Olzi}
\affiliation[e]{INFN, Sezione di Firenze, Via G. Sansone 1, I-50019 Sesto Fiorentino (Firenze), Italy}
\affiliation[f]{ Dipartimento di Fisica e Astronomia, Universit\'a di Firenze, Via G. Sansone 1, I-50019 Sesto Fiorentino (Firenze), Italy}
\emailAdd{andrea.olzi@unifi.it}

\abstract{
We revisit the holographic description of heavy–light mesons in the D3/D7 system at zero temperature, analyzing the dynamics of the coupled probe D7-branes through the non-Abelian Dirac--Born--Infeld action. Distinct quark masses are realized by separating the flavor branes, producing holographic flavor hierarchies. We refine the calculation made in previous works: we impose Hermiticity on the induced metric and fix the expansion of the determinant for matrix-valued fields. Implementing these improvements yields modified fluctuation equations and quantitatively different meson spectra: the scalar modes become heavier while the vector modes become lighter, removing the degeneracy reported in the literature. At finite 't Hooft coupling, we also observe a qualitatively different dependence of the vector modes on the quark masses. The resulting prescription provides a consistent, broadly applicable framework for incorporating non-Abelian flavor dynamics into holographic models and can be readily extended to situations away from the vacuum.
}

\maketitle
\flushbottom
\setcounter{page}{2}

\newpage

\section{Introduction}\label{sec:intro}

A key feature of the Standard Model is its flavor structure, characterized by mass hierarchies both between and within the three families of quarks and leptons. These hierarchies have wide-ranging consequences, and within quantum chromodynamics (QCD), they give rise to a spectrum of hadron masses that depends sensitively on the quark content. Any phenomenological model aiming to reproduce QCD physics accurately should therefore incorporate distinct quark masses for each flavor.

Flavors in holographic models were first introduced by adding probe D7-branes in an $AdS_5\times S^5$ geometry~\cite{Karch:2002sh}. This construction is dual to a low-energy effective theory living at a D3-D7 intersection and consisting of fields of ${\cal N}=4$ super Yang-Mills (SYM) theory coupled to hypermultiplets in the fundamental representation. The total supersymmetry is ${\cal N}=2$, and the number of D3-branes, $\Nc$, and D7-branes, $\Nf$, determine the rank of the gauge and flavor groups, respectively. In flat spacetime, the D3- and D7-branes intersect along $3+1$ directions, and there is a two-dimensional plane transverse to both types of branes, as indicated in table \ref{tab:D3D7} below.
\begin{table}[h!]
$$
\begin{array}{r|cccccccccc}
 & 0 & 1 & 2 & 3 & 4 & 5 & 6 & 7 & 8 & 9\\ \hline
\text{D3} & - & - & - & -& \bullet & \bullet & \bullet & \bullet & \bullet & \bullet \\
\text{D7} & - & - & - & -& - & - & - & - & \bullet & \bullet \\
\end{array}
$$
\caption{Brane configuration for the D3-D7 intersection. A horizontal line $(-)$ indicates a direction along which the brane is extended, while a bullet $(\bullet)$ denotes a direction in which the brane is localized.}\label{tab:D3D7}
\end{table}

Separating the D3- and D7-branes in this plane introduces masses for the flavors, since the open strings extending between them have a non-vanishing minimal length. Placing the D7-branes at different distances from the D3-branes allows us to realize distinct quark masses for each flavor, achieving our goal of introducing flavor-dependent mass hierarchies. In the holographic dual description, this translates into different asymptotic boundary conditions for the D7-brane embeddings in the background geometry, which result in different profiles in the bulk.

The D3-D7 model is obviously different from QCD, and the usual conditions of large-$\Nc$ and large 't Hooft coupling that lead to a semiclassical, small-curvature regime of the string theory dual apply. Nonetheless, the D3-D7 model can be extrapolated to construct phenomenological approximations to QCD, in particular to describe quenched flavors in a quark-gluon plasma~\cite{Kruczenski:2003be} or unpaired quark matter at large baryon density~\cite{Kobayashi:2006sb}, where supersymmetry might be less relevant. The same can be said of other string theory models with probe flavor branes, such as the Witten--Sakai--Sugimoto model~\cite{Sakai:2004cn}. Furthermore, the effective action describing the embedding of probe D7-branes in holographic setups has inspired the construction of bottom-up models aiming for a more realistic description of QCD, a prominent example being the V-QCD model~\cite{Jarvinen:2011qe}.

The effective action of the probe D7-branes is the well-known Dirac--Born--Infeld (DBI) action~\cite{Leigh:1989jq}. The meson spectrum can be obtained from solutions to the equations of motion derived from this action. This is also the spectrum of open strings with endpoints attached to the branes. If we have two stacks of probe branes corresponding to different quark masses, the meson spectrum splits into heavy-heavy, heavy-light, and light-light modes, depending on which brane the endpoints of open strings are attached. From the point of view of the DBI description, the functions that describe the embedding are promoted to matrices, with diagonal entries corresponding to heavy-heavy or light-light excitations, and the off-diagonal components to heavy-light excitations. Therefore, when considering different quark masses, one should use the full non-Abelian version of the DBI action, whose form was fixed by Myers using T-duality arguments \cite{Myers:1999ps}. 

The heavy–light meson spectrum in the D3/D7 model has been investigated previously using the non-Abelian DBI action~\cite{Erdmenger:2007vj} and, in the regime of large meson excitations, through classical open strings described by the Nambu–-Goto action~\cite{Erdmenger:2006bg}. In this work, we revisit the analysis based on the non-Abelian DBI action and identify two technical aspects that were not considered in earlier studies, both of which lead to modified fluctuation equations and consequently to different meson spectra:
\begin{itemize}
\item In the computation of the induced metric, an ordering ambiguity arises between the background metric and the derivatives of the embedding when generalizing from the Abelian to the non-Abelian case. The appropriate prescription is to define the components of the induced metric as  Hermitian matrices in flavor space, consistent with the fact that the brane embedding coordinates, and hence the quark positions, are real, as discussed above. This Hermiticity ensures that the action remains real and that the determinant and square root in the DBI action are well defined, since they act on matrices with real eigenvalues. In previous analyses, the Abelian ordering was adopted, breaking Hermiticity and leading to a complex induced metric in the non-Abelian case. 
\item To compute the heavy–light meson spectrum, we expand the non-Abelian DBI action to quadratic order in the fluctuations. This requires expanding a determinant over spacetime indices. 
Standard manipulations of determinants are often used for such expansions, but they assume that the factors inside the determinant commute. When these factors are themselves matrices, in our case, in flavor space, this assumption fails. We therefore perform the expansion using \textit{mixed-determinant formulas} that remain valid for non-commuting matrix-valued factors.
\end{itemize}

Properly accounting for the issues discussed above leads to fluctuation equations that differ from those obtained in earlier studies, and consequently to a quantitatively modified meson spectrum. The changes are most pronounced for the vector modes, where qualitative differences also appear. Our formulation reproduces the familiar Abelian equations and spectra in the limit of coincident flavor branes. Moreover, when the quark masses have equal magnitude but opposite signs, corresponding to the antipodal D7-brane configuration relative to the D3-branes, we recover complete agreement with previous results. This is consistent with the fact that in this configuration the mass-squared matrix becomes proportional to the identity, simplifying the determinant expansion.

The resulting spectra are shown in figures \ref{fig:scalar} and \ref{fig:vector} for the scalar and transverse vector modes, respectively, together with earlier results for comparison. In general, the scalar modes become heavier and the vector modes lighter when the corrected non-Abelian action is used, thereby lifting the degeneracy found in previous analyses. In addition, as displayed in figure \ref{fig:largeheavyquarkmass}, the lightest vector mode approaches zero mass when the heavy-quark mass becomes arbitrarily large.

The meson spectra are computed at several finite values of the ’t Hooft coupling. For the lowest vector modes, the dependence on the quark masses differs qualitatively from the expected infinite-coupling behavior. In that limit, the leading contribution to the meson mass corresponds to the energy of a classical string stretched between the two flavor branes, which equals the quark-mass difference and makes heavy–light mesons much heavier than their heavy–heavy or light–light counterparts. We find that while the scalar modes follow this expectation, the vector meson mass remains below the classical-string value even at large quark-mass separations. Nevertheless, the relative mass of the heavy–light states increases with the ’t Hooft coupling for both scalar and vector modes, suggesting that at sufficiently strong coupling the heavy–light mesons become heavier than the other modes, at least over part of the heavy-quark-mass range.

The content of the paper is as follows. In section \ref{sec:setup} we introduce the holographic background geometry and flavor branes, and expand the non-Abelian DBI action to the second order in fluctuations. In section \ref{sec:spectra} we derive the fluctuation equations and compute the meson spectra numerically, contrasting the results with analytic expressions obtained in the limits of small quark-mass difference and strong coupling. Section \ref{sec:discussion} summarizes our findings and outlines possible future directions. Technical details of the numerical analysis are collected in the appendices.

\section{Holographic setup}\label{sec:setup}

The holographic dual geometry to the ${\cal N}=4$ SYM theory with $SU(\Nc)$ gauge group is an $AdS_5\times S^5$ geometry with $\Nc$ units of five-form flux. The radius $L$ of $AdS$ and the sphere is the same, and it is related to the 't Hooft coupling $\lambda=\gYM^2 \Nc$ of the field theory through
\begin{equation}\label{eq:Ldef}
    L^4 = 4\pi \gs \Nc \ls^4 = 4\pi\lambda \ls^4 \ ,
\end{equation}
where $\ls=\sqrt{\alpha'}$ is the string length, and $\gs$ the string coupling. In the last equality, we have used $ \gs=\gYM^2$.

The $AdS_5\times S^5$ metric in Poincar\'e coordinates is
\begin{equation}
\dd s_{10}^2= g_{MN} \dd x^M \dd x^N=\frac{r^2}{L^2}\eta_{\mu\nu}\dd x^\mu \dd x^\nu+\frac{L^2}{r^2}\dd r^2 +L^2 \dd \Omega_5 \ ,
\end{equation}
with $M,N =0,\ldots,9$ and $\eta$ is assumed mostly plus. Here $x^\mu$, $\mu=0,1,2,3$ are the Minkowski field theory directions and $r$ the holographic radial direction, with the asymptotic boundary of $AdS$ at $r\to \infty$. It will be convenient to group the $S^5$ part of the metric with the $AdS$ radial coordinate to have a conformally $\mathbb{R}^6$ factor in the geometry, which we then divide into $\mathbb{R}^4\times \mathbb{R}^2$:
\begin{equation}
\dd r^2+r^2 \dd \Omega_5^2=\dd \rho^2+\rho^2\dd \Omega_3+\dd y^2+\dd z^2\ ,\quad r^2=\rho^2+y^2+z^2\ .
\end{equation}
Here $x^4=\rho$ is a spherical radial coordinate in $\mathbb{R}^4$ and $x^8=y$, $x^9=z$ are Cartesian coordinates. The metric for the unit radius $S^3$ can be written as
\begin{equation}
    \dd \Omega_3=\gamma_{\alpha\beta}^{S^3}\dd \theta^\alpha \dd \theta^\beta \ ,
\end{equation}
where we have introduced a set of angular coordinates $\theta^\alpha,\theta^\beta$, with $\alpha,\beta=5,6,7$. In these coordinates, the ten-dimensional metric becomes
 \begin{equation}\label{eq:metric10d}
\dd s_{10}^2=\frac{\rho^2+y^2+z^2}{L^2}\eta_{\mu\nu}\dd x^\mu \dd x^\nu+\frac{L^2}{\rho^2+y^2+z^2}\left( \dd \rho^2+\rho^2 \dd \Omega_3+\dd y^2+\dd z^2\right)\ .
\end{equation}
The plane spanned by the $y,z$ coordinates can be identified as the space orthogonal to both D3- and D7-branes. A general embedding of the flavor D7-branes will be extended along the field theory directions $x^\mu$, the $S^3$ part, and it will describe a curve in the space spanned by the $(\rho,y,z)$ coordinates. For a single D7-brane, we can select $z=0$ and describe the embedding through a profile function $y(\rho)$ without loss of generality. It can be shown, see \cite{Karch:2002sh}, that the general solution for the embedding of a D7-brane in the $AdS_5\times S^5$ background, for $z=0$, is given by $y(\rho)=$ constant. For multiple D7-branes, we keep the brane embeddings at $z=0$ for simplicity. The profile in the $y$-direction is generalized to a diagonal $\Nf\times \Nf$ matrix consisting of two diagonal blocks, each proportional to the identity matrix $\mathbb{1}_{\Nf/2}$. In the most general case, the embeddings in both the $y$ and $z$ directions could be nontrivial matrices and need not be diagonal. However, for describing two sets of flavors with different masses, the block-diagonal configuration is sufficiently general. In practice, we omit the explicit identity matrices in each block, so some of the resulting formulas take the same form as in the $\Nf=2$ case.

\subsection{The D7-brane action}

As explained in the introduction, we have to deal with non-Abelian configurations of the D7-brane fields. Denoting by $\sigma^a$, $a=0,\dots,7$ the coordinates on the worldvolume of the D7-branes, the independent bosonic fields living on the D7-brane are the $SU(\Nf)$ gauge fields $A_a(\sigma)$ and the embedding functions along the transverse directions $X^I(\sigma)$, $I=8,9$, which are in the adjoint representation of the D7-brane gauge group. Throughout the text, we will use uppercase letters for objects that are matrices with flavor indices, and lowercase letters for the components of the matrices and scalar quantities. We choose the embeddings such that $X^a=\sigma^a \mathbb{1}_{\Nf}$. We adopt the same index conventions for the background coordinates $x^M$: the indices $a,b$ label directions parallel to the D7-branes, and the indices $I,J$ denote directions transverse to the branes.

Denoting by $P[\cdot]$ the pullback to the D7-brane worldvolume, the D7-brane action depends on the pullback of the background dilaton $\Phi=P[\phi]$, and on the induced metric and the Kalb--Ramond field through the combination
\begin{equation}
E_{MN}=P[g_{MN}+b_{MN}] \ .
\end{equation}
In the background we consider, the dilaton is constant, and the Kalb--Ramond field vanishes, so neither contributes to the action or the spectrum. 

Then, for an Abelian configuration, the pullback of the metric takes the form
\begin{equation}
E_{ab}=P[g_{ab}]=G_{MN}(X) \partial_a X^M\partial_b X^N=\left(g_{ab}+g_{IJ}\partial_a x^I\partial_b x^J \right)\bigg|_{y=x^8(\rho),z=x^9(\rho)}\mathbb{1}_{\Nf} \ .
\end{equation}
When promoting this expression to a non-Abelian configuration, there is an ambiguity in how we order the metric factor with the derivatives. Since the other terms that are combined with the induced metric are Hermitian, and the action should depend on Hermitian fields in order to be real in every case, this strongly suggests that $G_{ab}$  should be Hermitian, in which case the generalization to a non-Abelian configuration must be
\begin{equation}
E_{ab}=P[g_{ab}]=G_{ab}(X)+\frac{1}{2}\left(D_a X^I G_{IJ}(X) D_b X^J+D_b X^I G_{IJ}(X) D_a X^J\right) \ ,
\end{equation}
where the partial derivatives have to be promoted to covariant derivatives
\begin{equation}
    D_a X^I=\partial_a X^I+i[A_a,X^I]\ .
\end{equation}
It turns out that for a non-Abelian configuration, one should introduce an additional quantity,
\begin{equation}
Q^I_{\ J}=\delta^I_{\ J}\mathbb{1}_{\Nf}+\dfrac{1}{4\pi \ls^2}\{i[X^I,X^K]\,,\, E_{KJ}\} \ .
\end{equation}
Compared to the expression that is usually presented, we have introduced an anticommutator in the second term in order to make $Q^I_{\ J}$ a Hermitian matrix of flavor indices.
 
Taking all this into account, Myers' action is \cite{Myers:1999ps}
\begin{equation}\label{eq:actionmeyers}
S_\text{D7}=-T_7\int \dd ^8\sigma\,\text{STr}_{\Nf} \left[e^{-\Phi}\sqrt{-\det\left(E_{ab}+E_{aI}(Q^{-1}-\delta \mathbb{1}_{\Nf})^{IJ}E_{Jb}+2\pi \ls^2 F_{ab}\right) \det Q^I_{\ J}} \right] \ ,
\end{equation}
where $T_7=1/((2\pi)^7 \gs \ls^8)$ is the D7-brane tension and $F_{ab}=\partial_a A_b-\partial_b A_a+i[A_a,A_b]$ the field strength of the gauge fields. The determinant $\det(E_{ab}+\cdots)$ inside the square root is taken over the worldvolume indices $a,b$ while $\det Q^I_{\ J}$ is taken over the transverse indices $I,J$. The symmetrized trace is taken over the $SU(\Nf)$ flavor indices.

From now on, we will absorb the $2\pi \ls^2$ factors in the fields and coordinates and work with dimensionless variables. In particular,
\begin{align}\label{eq:coordtransf}
    &X^I\to \sqrt{2\pi}\, \ls\,X^I\ ,&  &L\to \sqrt{2\pi} \,\ls\,L\ ,& &A_a\to (\sqrt{2\pi}\, \ls)^{-1}\,A_a.&\sigma^a\to \sqrt{2\pi}\, \ls\,\sigma^a \ .&
\end{align}
As a consequence, using \eqref{eq:Ldef}, the rescaled $L$ reads $(\lambda/\pi)^{1/4}$.

It should be noted that Myers' action with the symmetrized trace prescription is known to be correct up to quartic terms in the gauge fields or scalar commutators, but appears ambiguous for higher-order terms \cite{Tseytlin:1997csa,Hashimoto:1997gm,Myers:1999ps}, although there are concrete proposals for additional derivative contributions up to ${\mathcal{O}}((\alpha')^4)$ \cite{Koerber:2002zb,Keurentjes:2004tu}. For the calculation of the spectrum we aim at, the Myers action is sufficient.

The total action for the D7-branes also involves the Wess--Zumino term, which reads \cite{Myers:1999ps}
\begin{equation}\label{eq:WZD7}
    S_\text{D7}^\text{WZ} = T_7\dfrac{(2\pi \ls^2)}{2}\int \text{STr}_{\Nf}\left(P[e^{i\,2\pi\ls^2\iota_X\iota_X}C_4] \wedge \left(F\wedge F\right)\right) \ ,
\end{equation}
where $C_4$ is the Ramond--Ramond four-form, and $\iota_X$ denotes the interior product with the transverse matrix scalars. The interior product acts on forms as an anticommuting operator of form degree -1.
We will choose a configuration such that this term does not contribute, and we will neglect it for the rest of this work.

\subsection{Background solution and expansions to quadratic order in the fluctuations}

Since we are considering a diagonal background for the D7-brane fields, the solutions can be obtained from the Abelian action and are known to be \cite{Karch:2002sh}
\begin{equation}
X^8=M=\left(\begin{array}{cc} m_1 &  0\\0  & m_2\end{array}\right)\ ,\quad  X^9=0\ ,\quad A_a=0 \ ,
\end{equation}
where, as already pointed out, $m_1$ and $m_2$ are constant functions in $\rho$ describing the embedding of the D7-branes in the background geometry. Note that, in order to have the D7-branes separated from the D3-branes a distance of the order of the $AdS$ radius $L$, we should require $m_{1,2}\sim\lambda^{1/4}$.

If we restore the proper units, the quark masses $\mathbf{m_{q}}_i$, $i=1,2$, are obtained from the value of the $X^8$ field through the equation
\begin{equation}\label{eq:quarkmassdef}
m_i=\sqrt{2\pi}\, \ls \mathbf{m_{q}}_i \ ,
\end{equation}
where we recall that $m_i$ are the dimensionless locations of the D7-brane along the $X^8$ direction.

\begin{figure}
\center
\includegraphics[width = 0.65\textwidth]{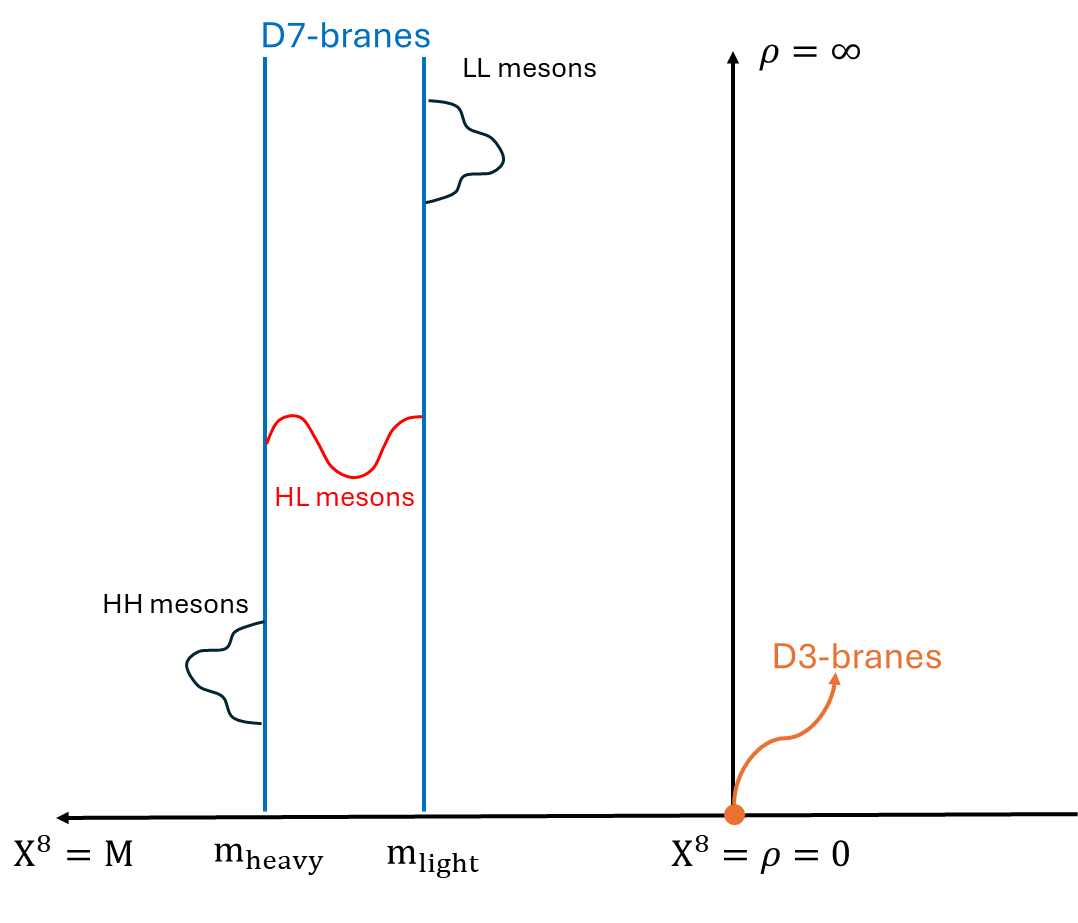}\caption{D7-branes' embedding in the $(\rho,X^8)$-subspace. They are placed at $X^9=0$ and their position along the $X^8$ direction, $m_\text{heavy}$ and $m_\text{light}$, is constant in $\rho$.  Strings stretching between the same D7-branes correspond to light-light (LL) and heavy-heavy (HH) mesonic excitations. String stretching between flavor branes placed at different positions in the $X^8$ corresponds to the heavy-light (HL) mesons.}
\label{fig:setup}
\end{figure}

In figure \ref{fig:setup}, the brane setup is schematically presented in the $(\rho,X^8)$-subspace. The locations of the D7-branes, along the $X^8$ direction, are named $m_\text{heavy}$ and $m_\text{light}$ to emphasize that the more the brane is distant from the $X^8=0$ axis, the more massive is the dual quark operator. In general, the positions of the branes will be referred to as $m_1$ and $m_2$.

We now introduce fluctuations around the background solutions
\begin{equation}
X^8=M+Y\ ,\quad  X^9=Z \ ,\quad  A_a \ne 0 \ ,
\end{equation}
with components $y_{ij}$, $z_{ij}$, and $(a_a)_{ij}$.\footnote{Following this notation, the non-zero matrix elements of $M$ are $m_{ii}=m_i$ for $i=1,2$.} In order to compute the spectrum, it is enough to consider fluctuations at quadratic order in the action, in which case diagonal and off-diagonal fluctuations are decoupled. Since we are interested in the heavy-light meson spectrum, the only non-zero components will be those with $i\neq j$.  We will also restrict ourselves to fluctuations that do not depend on the $S^3$ directions,\footnote{This corresponds to consider mesons which do not carry angular momentum along the three-sphere.} so that $A_\alpha=0$ and $Y=Y(x^\mu,\rho)$, $Z=Z(x^\mu,\rho)$, $A_\mu=A_\mu(x^\mu,\rho)$, $A_\rho=A_\rho(x^\mu,\rho)$, although we will keep the formulas as general as possible in the intermediate steps. As a consequence of this choice, the Wess--Zumino term \eqref{eq:WZD7} vanishes at quadratic order.

Let us now go over some simplifications. Note first that the derivatives and the field strength of the background fields vanish, since they are constant, and that covariant derivatives and field strengths, involving the fluctuations, appear at least quadratically in the action, so we need to keep only the part that is linear in the fluctuations, namely
\begin{equation}
 D_a Y\sim \partial_a Y +i[A_a,M]\ ,\quad D_a Z\sim \partial_a Z\ ,\quad F_{ab}\sim \partial_a A_b-\partial_b A_a\ .
\end{equation}
Another simplification comes from observing that in our background geometry $E_{MN}=G_{MN}$ is diagonal in the $M,N$ indices, so that one must have that $P\left[E_{aI}(Q^{-1}-\delta \mathbb{1}_{\Nf})^{IJ}E_{Jb}\right]$ is quadratic in the fluctuations through the derivatives $D_a X^I$. On the other hand, since the background is diagonal in flavor indices, $Q^{-1}-\delta \mathbb{1}_{\Nf}$ should be at least linear in the fluctuations. Thus, the corresponding terms in the action do not contribute to quadratic order.

We move on now to the expansion of the induced metric to quadratic order
\begin{equation}
G_{ab}\sim G^{(0)}_{ab}+G_{ab}^{(1)}+G_{ab}^{(2)}\ ,
\end{equation} 
where the background value of any metric component is
\begin{equation}
G^{(0)}_{MN}=G_{MN}(X^8=M,X^9=0) \ .
\end{equation}
There are two types of terms in the induced metric, ones originating in the expansion of $G_{ab}$,
\begin{equation}
G_{ab}(X)\sim G^{(0)}_{ab}+H_{ab}^{(1)}+H_{ab}^{(2)}\ ,
\end{equation}
and the others from terms containing derivatives. The linear contribution to the induced metric is fully determined by the metric expansion $G_{ab}^{(1)}=H_{ab}^{(1)}$, while the quadratic contribution is
\begin{equation}
G_{ab}^{(2)}=H_{ab}^{(2)}+\frac{1}{2}\left(D_a X^I G^{(0)}_{IJ} D_b X^J+ D_b X^I G^{(0)}_{IJ} D_a X^J\right)\ .
\end{equation}
Since $G^{(0)}_{IJ}$ depends only on the background embedding, it is diagonal in flavor space, whereas $D_a X^I$ has only off-diagonal components. As a result, the product $D_a X^I G^{(0)}_{IJ} D_b X^J$ is diagonal, with components
\begin{equation}
\left(D_a X^I G^{(0)}_{IJ} D_b X^J\right)_{ii}=(D_a X^I)_{ij}\left(G^{(0)}_{IJ}\right)_{jj}(D_b X^J)_{ji} \ .
\end{equation}
The full formulas for the pullback metric are, for the field theory directions,
\begin{equation}
P[E_{\mu\nu}]=P[G_{\mu\nu}]=G_\text{Mink}\eta_{\mu\nu}=\sqrt{\frac{\pi}{\lambda}}\begin{pmatrix}
\rho^2+m_1^2 +y_{12}y_{21} & (m_1+m_2)y_{12} \\
(m_1+m_2)y_{21} & \rho^2+m_2^2 +y_{12}y_{21} 
\end{pmatrix}\eta_{\mu\nu} \ .
\end{equation}
We recall that, according to the rescalings \eqref{eq:coordtransf}, the $AdS$ radius reads $L = (\lambda/\pi)^{1/4}$. In the directions along the $S^3$ one gets
\begin{equation}
P[E_{\alpha\beta}]=P[G_{\alpha\beta}]=G_{\rho\rho} \rho^2 \gamma_{\alpha\beta}^{S^3}=\sqrt{\frac{\lambda}{\pi}} \dfrac{1}{d} \begin{pmatrix}
\rho^2+m_2^2 +y_{12}y_{21} & -(m_1+m_2)y_{12} \\
-(m_1+m_2)y_{21} & \rho^2+m_1^2 +y_{12}y_{21} 
\end{pmatrix}\rho^2 \gamma_{\alpha\beta}^{S^3}\ ,
\end{equation}
where $\gamma_{\alpha\beta}^{S^3}$ is the metric of the unit round $S^3$ and
\begin{equation}
    d = (m_1 m_2 - 
   y_{12} y_{21})^2 + (m_1^2 + m_2^2 + 2 y_{12} y_{21}) \rho^2 + \rho^4 \ .
\end{equation}
Note that $G_\text{Mink}G_{\rho\rho}=\mathbb{1}_{\Nf}$. We can find $H^{(1)}_{ab}$ and $H^{(2)}_{ab}$ by direct expansion in the fluctuations of the formulas above. However, for a larger number of flavors or more complicated metrics, it would be useful to be able to find the terms in the expansion more systematically. We provide here expressions that can be applied to any metric.
The metric evaluated on the non-Abelian embedding is computed through a Taylor expansion
\begin{align}
    G_{MN}(X^8,X^9) = \sum\limits_{n=0}^\infty\sum_{l=0}^\infty\dfrac{1}{n!\ l!}\dfrac{\partial^n}{\partial y^n}\dfrac{\partial^l}{\partial z^l}g_{MN}(y,z)\bigg|_{y=z=0}\mathcal{S}[(X^8)^n(X^9)^l] \ .
\end{align}
where $\mathcal{S}[(X^8)^n(X^9)^l]$ is the symmetrization operation, \emph{i.e.}, is the average over all distinct products made by placing $n$ copies of $X^8$ and $l$ copies of $X^9$ in every possible order.

We find
\begin{align}
    &G^{(0)}_{MN}=G_{MN}(M,0)=\sum\limits_{n=0}^\infty\dfrac{1}{n!}\dfrac{\partial^n}{\partial y^n}g_{MN}\bigg|_{y=z=0}M^n \\
    &H^{(1)}_{MN} = \sum\limits_{n=0}^\infty\dfrac{1}{n!}\dfrac{\partial^n}{\partial y^n}g_{MN}\bigg|_{y=z=0}\sum\limits_{k=0}^{n-1}M^k Y\,M^{n-1-k} \\ \nonumber
    &H^{(2)}_{MN}  = \sum\limits_{n=0}^\infty\dfrac{1}{n!}\dfrac{\partial^n}{\partial y^n}g_{MN}\bigg|_{y=z=0}\sum\limits_{k=0}^{n-2}\sum\limits_{r=0}^{k}M^{k-r}Y M^r Y M^{r-k-2}+ \sum\limits_{n=0}^\infty\dfrac{1}{2\, n!}\dfrac{\partial^n}{\partial y^n}\dfrac{\partial^2}{\partial z^2}g_{MN}\bigg|_{y=z=0}M^n Z^2  \\
    &\quad\hspace{0.6cm} =\sum\limits_{n=0}^\infty\dfrac{1}{n!}\dfrac{\partial^n}{\partial y^n}g_{MN}\bigg|_{y=z=0}\sum\limits_{k=0}^{n-2}\sum\limits_{r=0}^{k}M^{k-r} Y M^r Y M^{r-k-2} + \dfrac{1}{2}\dfrac{\partial^2}{\partial z^2}g_{MN}\bigg|_{y=z=0} Z^2 \ .
\end{align}
We will now use the following resummations for a diagonal matrix $D$, with components $d_{ii}$, and a matrix $X$ with only off-diagonal components $x_{ij}$, $i\neq j$. We use the indices $i,j$ to keep track of diagonal and off-diagonal flavor components; they label any pair of flavors (or flavor blocks) with different masses. In the expressions below, repeated $i,j$ indices do {\bf not} imply summation, while Einstein summation is maintained for all other indices. We find
\begin{align}
&\sum_{k=0}^{n-1} (D^k X D^{n-1-k})_{ij} = \sum_{k=0}^{n-1} (d_{ii})^k(d_{jj})^{n-1-k} x_{ij}=\frac{(d_{ii})^n-(d_{jj})^n}{d_{ii}-d_{jj}}x_{ij} \\
&\sum_{k=0}^{n-2} \sum_{l=0}^k (D^{k-l}X D^l X D^{n-2-k})_{ii}=\sum_{k=0}^{n-2} \sum_{l=0}^k (d_{ii})^{n-2-l}(d_{jj})^{l} x_{ij}x_{ji} \notag\\
&\hspace{4.5cm}\quad\, =\left[n\,\dfrac{(d_{ii})^{n-1}}{d_{ii}-d_{jj}}- \dfrac{(d_{ii})^n-(d_{jj})^n}{(d_{ii}-d_{jj})^2}\right]x_{ij}x_{ji}\,.
\end{align}
Using the formulas above, we arrive at the following expressions for the non-zero metric components
\begin{align}
    &(h^{(1)}_{MN})_{ij}= \dfrac{g_{MN}(m_i,0)-g_{MN}(m_j,0)}{m_i-m_j}y_{ij} \\
    &(h^{(1)}_{MN})_{ji}= \dfrac{g_{MN}(m_i,0)-g_{MN}(m_j,0)}{m_i-m_j}y_{ji} \\
        &(h^{(2)}_{MN})_{ii} = \left(\dfrac{\partial_y g_{MN}(m_i,0)}{m_i-m_j}-\dfrac{g_{MN}(m_i,0)-g_{MN}(m_j,0)}{(m_i-m_j)^2}\right)y_{ij}y_{ji}+ \dfrac{1}{2}\partial_z^2g_{MN}(m_i,0)z_{ij}z_{ji} \\
         &(h^{(2)}_{MN})_{jj} = \left(\dfrac{\partial_yg_{MN}(m_j,0)}{m_j-m_i}-\dfrac{g_{MN}(m_j,0)-g_{MN}(m_i,0)}{(m_i-m_j)^2}\right)y_{ij}y_{ji}+ \dfrac{1}{2}\partial_z^2g_{MN}(m_j,0)z_{ij}z_{ji}\ .
\end{align}
We have checked that the formulas above produce the same result as the direct calculation.

Finally, the result for the terms in the expansion of the metric is
\begin{align}
G^{(0)}_{\mu\nu} &=\sqrt{\frac{\pi}{\lambda}}\begin{pmatrix}
\rho^2+m_1^2  & 0 \\
0 & \rho^2+m_2^2  
\end{pmatrix}\eta_{\mu\nu} \\
G^{(0)}_{\alpha\beta} &=\sqrt{\frac{\lambda}{\pi}}\begin{pmatrix}
(\rho^2+m_1^2)^{-1}  & 0 \\
0 & (\rho^2+m_2^2)^{-1}  
\end{pmatrix} \rho^2 \gamma_{\alpha\beta}^{S^3} \\
H_{\mu\nu}^{(1)} &= \sqrt{\frac{\pi}{\lambda}}(m_1+m_2)\begin{pmatrix}
0 & y_{12} \\
y_{21} & 0 
\end{pmatrix}\eta_{\mu\nu} \\
H_{\alpha\beta}^{(1)} &=- \sqrt{\frac{\lambda}{\pi}}\dfrac{m_1+m_2}{(\rho^2+m_1^2)(\rho^2+m_2^2)}\begin{pmatrix}
0 & y_{12} \\
y_{21} & 0 
\end{pmatrix} \rho^2 \gamma_{\alpha\beta}^{S^3} \\
H_{\mu\nu}^{(2)} &=\sqrt{\frac{\pi}{\lambda}}\,\,\mathbb{1}_2 \, y_{12}y_{21}\, \eta_{\mu\nu}  \\
H_{\alpha\beta}^{(2)} &=\sqrt{\frac{\lambda}{\pi}}\begin{pmatrix}
\frac{2 m_1 m_2+m_1^2-\rho ^2}{\left(m_2^2+\rho ^2\right) \left(m_1^2+\rho ^2\right)^2} & 0 \\
0 &  \frac{2 m_1 m_2+m_2^2-\rho ^2}{\left(m_1^2+\rho ^2\right) \left(m_2^2+\rho ^2\right)^2}
\end{pmatrix}y_{12}y_{21} \rho^2 \gamma_{\alpha\beta}^{S^3} \ .
\end{align}
For the components transverse to the D7-brane the metric is proportional to the metric along the $S^3$ directions, thus
\begin{equation}
G^{(0)}_{IJ}=\sqrt{\frac{\lambda}{\pi}}\begin{pmatrix}
(\rho^2+m_1^2)^{-1}  & 0 \\
0 & (\rho^2+m_2^2)^{-1}  
\end{pmatrix} \delta_{IJ} \ .
\end{equation}

\subsection{Determinants to quadratic order in the fluctuations}

When expanding the DBI action \eqref{eq:actionmeyers} for Abelian configurations, it is common to separate background and fluctuation 
\begin{equation}
K_{ab} = K_{ab}^{(0)} + \Xi_{ab},
\end{equation}
where $K_{ab}$ are the components of a $d\times d$ matrix, with $K_{ab}^{(0)}$ the background and $\Xi_{ab}$ the fluctuation.

The contributions to the determinant are split according to
\begin{equation}\label{eq:naiveexp}
\det(K_{ab}^{(0)}+\Xi_{ab}) = \det (K_{ab}^{(0)})\,\det\!\left(\delta^{a}{}_{b} + ({K^{(0)}}^{-1})^{ac}\,\Xi_{cb}\right)\ .
\end{equation}
The second factor is then expanded using $\log\det=\operatorname{Tr}\log$, leading to
\begin{equation}
\det(\delta^{a}{}_{b} + ({K^{(0)}}^{-1})^{ac}\,\Xi_{cb})
  = \exp\!\left[\operatorname{Tr}\log(\delta^{a}{}_{b} + ({K^{(0)}}^{-1})^{ac}\,\Xi_{cb})\right] \ ,
\end{equation}
whose perturbative expansion yields
\begin{equation}
\begin{aligned}
\approx \delta^{a}{}_{a}
&+ ({K^{(0)}}^{-1})^{ac}\,\Xi_{ca} \\
&+ \frac12\!\left[
   \big(({K^{(0)}}^{-1})^{ac}\,\Xi_{ca}\big)^{2}
   - ({K^{(0)}}^{-1})^{ac}\,\Xi_{cb}
     ({K^{(0)}}^{-1})^{bd}\,\Xi_{da}
   \right]
+ \ldots \ .
\end{aligned}
\end{equation}
However, when $K^{(0)}_{ab}$ and $\Xi_{ab}$ are themselves matrices in flavor space, the above manipulations become ambiguous: the factors no longer commute, the splitting $\det(K^{(0)}+\Xi)=\det (K^{(0)})\,\det(\ldots)$ is not unique, and even the definition of the determinant requires an ordering prescription. Although the symmetrized trace is intended to resolve such ambiguities, it must be implemented at the level of the determinant itself.

A useful observation is that the determinant of a $d\times d$ matrix can be written in a manifestly symmetric form,
\begin{equation}
\det K_{ab}
  = \frac{1}{d!}\,\epsilon^{a_1\cdots a_d}\epsilon^{b_1\cdots b_d} K_{a_1 b_1}\cdots K_{a_d b_d} \ ,
\end{equation}
where all permutations of the matrix entries appear symmetrically. This expression generalizes straightforwardly to the non-Abelian case and is naturally compatible with the symmetrized trace prescription. The \textit{mixed-determinant} formulas that we employ below arise by expanding this fully symmetric determinant expression after splitting the matrix into background and perturbation, and keeping terms up to the required order in $\Xi_{ab}$.

Let us proceed to compute the simpler determinant $\det Q^I_{\ J}$ to quadratic order:
\bea
\det Q^I_{\ J} & = & \frac{1}{2}\epsilon_{I_1 I_2}\epsilon^{J_1 J_2}Q^{I_1}_{\ J_1}Q^{I_2}_{\ J_2}  \\
&= & \mathbb{1}_{\Nf}+ \dfrac{1}{2}\{i[X^I,X^K]\,,\, E_{KI}\}
 +\frac{1}{8}\epsilon_{I_1 I_2}\epsilon^{J_1 J_2}\{i[X^{I_1},X^{K_1}]\,,\, E_{K_1 J_1}\}\{i[X^{I_2},X^{K_2}]\,,\, E_{K_2 J_2}\} \nonumber  \ ,
\eea
where $\epsilon^{I_1I_2}$ is the two-dimensional Levi--Civita symbol. Since $E_{IJ}=G_{IJ}$ is symmetric in its indices, the second term vanishes. Using now $\epsilon_{I_1 I_2}\epsilon^{J_1 J_2}=\delta_{I_1}^{J_1}\delta^{J_2}_{I_2}-\delta_{I_2}^{J_1}\delta^{J_1}_{I_2}$, we arrive at
\begin{equation}
\det Q^I_{\ J}=\mathbb{1}_{\Nf}-\frac{1}{8}\{i[X^{I_1},X^{K_1}]\,,\, E_{K_1 I_2}\}\{i[X^{I_2},X^{K_2}]\,,\, E_{K_2 I_1}\} \ .
\end{equation}
The linear order contribution to the anticommutator is
\begin{equation}
\{i[X^{I},X^{K}]\,,\, E_{K J}\}\sim \left(\begin{array}{cc} 0 & \{ i[M,Z],G^{(0)}_{zz}\}  \\  \{ i[Z,M],G^{(0)}_{yy}\} & 0 \end{array}\right)=\left(\begin{array}{cc} 0 & \{ i[M,Z],G^{(0)}_{zz}\}  \\  \{ -i[M,Z],G^{(0)}_{zz}\} & 0 \end{array}\right) \ ,
\end{equation}
where we have used that $G_{yy}=G_{zz}$ for our background metric. Therefore,
 \begin{equation}
\det Q^I_{\ J}\sim \mathbb{1}_{\Nf}+Q^{(2)}\equiv  \mathbb{1}_{\Nf}-\frac{1}{4}\{ [Z,M],G^{(0)}_{zz}\}\{ [Z,M],G^{(0)}_{zz}\}\ .
\end{equation}
The non-zero flavor components of $Q^{(2)}$ are in the diagonal
\begin{equation}
    q^{(2)}_{ii}\sim  c_{ij}^{zz} z_{ij}z_{ji}\ ,
\end{equation}
where we have defined
\begin{equation}
    c_{ij}^{zz}=\frac{1}{4}(m_i-m_j)^2\left(g_{zz}(m_i,0)+g_{zz}(m_j,0)\right)^2=\frac{\lambda}{\pi}\frac{(m_i-m_j)^2}{(\rho^2+m_i^2)^2(\rho^2+m_j^2)^2}\left(\rho^2+\frac{m_i^2+m_j^2}{2}\right)^2\ .
\end{equation}
We observe that $c_{ij}^{zz} = c_{ji}^{zz}$ and therefore $q_{ii}^{(2)}=q_{jj}^{(2)}$.

Finally, we have to expand to quadratic order the more complicated determinant
\begin{equation}\label{eq:detK}
\det\left(E_{ab}+F_{ab}\right)\equiv \det K_{ab}=\frac{1}{8!}\epsilon^{a_0\cdots a_7}\epsilon^{b_0\cdots b_7} K_{a_0 b_0}\cdots K_{a_7 b_7} \ .
\end{equation} 
First, the matrix $K_{ab}$ has an expansion to quadratic order
\begin{equation}
    K_{ab}=K^{(0)}_{ab}+\Xi_{ab}\sim K^{(0)}_{ab}+\Xi_{ab}^{(1)}+\Xi_{ab}^{(2)}\ ,
\end{equation}
where the background value is just the background induced metric $K^{(0)}_{ab}=G^{(0)}_{ab}$, and 
\begin{equation}
    \Xi_{ab}^{(1)}=H_{ab}^{(1)}+F_{ab}\ ,\quad \Xi_{ab}^{(2)}=H_{ab}^{(2)}\ .
\end{equation}
The background and the term quadratic in the fluctuations have non-zero diagonal components $(k^{(0)}_{ab})_{ii}=(g^{(0)}_{ab})_{ii}$, $(\xi^{(2)}_{ab})_{ii}$, while the term linear in the fluctuation only has off-diagonal components  $(\xi^{(1)}_{ab})_{ij}$. 

Note that terms in \eqref{eq:detK} with a single factor of $F_{ab}$ would vanish, so we do not have to consider commutators $[A_a,A_b]$ to quadratic order. The expansion of the determinant in turn to quadratic order is
\begin{equation}
    \det K_{ab}\sim \Delta_0+\Delta_1^{(1)}+\Delta_1^{(2)}+\Delta_2^{(2)}\ ,
\end{equation}
where we have defined ($n=1,2$)
\begin{align}
&\Delta_0 = \det(K^{(0)}_{ab})=\det(G^{(0)}_{ab}) \\
&\Delta_1^{(n)} = \frac{1}{8!}\sum_{k=0}^7 \epsilon^{a_0\cdots a_k\cdots a_7}\epsilon^{b_0\cdots b_k \cdots  b_7} K^{(0)}_{a_0 b_0}\cdots K^{(0)}_{a_{k-1} b_{k-1}} \Xi_{a_k b_k}^{(n)} K^{(0)}_{a_{k+1} b_{k+1}}\cdots K^{(0)}_{a_7 b_7} \\ \nonumber
&\Delta_2^{(2)} = \frac{1}{8!}\sum_{k=1}^7\sum_{l=0}^{k-1} \epsilon^{a_0\cdots a_l\cdots  a_k\cdots a_7}\epsilon^{b_0\cdots b_l\cdots b_k \cdots  b_7} K^{(0)}_{a_0 b_0} \\
&\hspace{3cm}\,\,\,\,\cdots K^{(0)}_{a_{l-1} b_{l-1}} \Xi_{a_l b_l}^{(1)} K^{(0)}_{a_{l+1} b_{l+1}} \cdots K^{(0)}_{a_{k-1} b_{k-1}} \Xi_{a_k b_k}^{(1)} K^{(0)}_{a_{k+1} b_{k+1}}\cdots K^{(0)}_{a_7 b_7}\ ,
\end{align}
which correspond respectively to zeroth-, first-, and second-order contribution in the expansion in $\Xi$ of $\det K_{ab}$, respectively. 

In order to compute efficiently the terms appearing in the expansion of the determinant we use the following identity \cite{2013arXiv1306.1315A,2018arXiv180605105B}: given $A_1,\ldots,A_n$ an $n$-tuple of $n\times n$ matrices, with elements $a_{ij}^k$, $k = 1,\ldots,n$, the \textit{mixed determinant} is defined as
\begin{equation}\label{eq:mixeddet}
    D(A_1,\ldots,A_n) = \frac{1}{n!}\sum\limits_{\sigma,\tau\in S_n}\text{sign}(\sigma)\prod\limits_{i=1}^n a_{i\,\sigma(i)}^{\tau(i)} = \frac{1}{n!}\dfrac{\partial^n}{\partial t_1 \cdots\partial t_n}\det(t_1A_1+\ldots+t_nA_n)\bigg|_{t_1=\cdots=t_n=0},
    \end{equation}
where $S_n$ is the symmetric group of permutations $\{1,\ldots,n\}$. If $A_k = A$ for all $k$, this formula reduces to the standard determinant. Taking into account that $K^{(0)}_{ab}$ and $\Xi^{(2)}$ are diagonal matrices in flavor and $\Xi^{(1)}$ has only off-diagonal components, the non-zero flavor components of the determinants are
\begin{align}
&(\Delta_0)_{ii} = \det[g^{(0)}_{ii}]\ ,\\
&(\Delta_1^{(1)})_{ij} =\frac{1}{8!}\sum_{k=0}^7 \frac{\partial^8}{\partial t_0\cdots\partial t_7}\det\bigg[\sum_{n_i<k} t_{n_i}k^{(0)}_{ii}+t_k\xi^{(1)}_{ij}+\sum_{n_j>k} t_{n_j}k^{(0)}_{jj} \bigg] \ ,\\ 
&(\Delta_1^{(2)})_{ii} = \frac{1}{8!}\sum_{k=0}^7 \frac{\partial^8}{\partial t_0\cdots\partial t_7}\det\bigg[\sum_{n_i<k} t_{n_i}k^{(0)}_{ii}+t_k\xi^{(2)}_{ii}+ \sum_{n_i'>k} t_{n_i'}k^{(0)}_{ii}\bigg] \ ,\\
&(\Delta_2^{(2)})_{ii} =\frac{1}{8!}\sum_{k=0}^6\sum_{l=k+1}^7 \frac{\partial^8}{\partial t_0\cdots\partial t_7}\det\bigg[\sum_{n_i<k} t_{n_i}k^{(0)}_{ii}+t_k\xi^{(1)}_{ij} 
+\sum_{k<n_j<l} t_{n_j} k^{(0)}_{jj} +t_l \xi^{(1)}_{ji}+\sum_{n_i'>l} t_{n_i'}k^{(0)}_{ii}\bigg] \ .
\end{align}
Here, we have omitted the spacetime indices $ab$ over which the determinants are taken.\\
Using now that the fluctuations do not depend on the $S^3$ directions, we arrive at
\begin{align}
&(\Delta_0)_{ii}= -\gamma^{S^3}\rho^6 \\
&(\Delta_1^{(1)})_{ij} = 0 \\
&(\Delta_1^{(2)})_{ii} = -\gamma^{S^3}\rho^6\left(c^{xx}_{ij}\eta^{\mu\nu}\delta_{IJ}D_\mu x^I_{ij}D_\nu x^J_{ji}+c^{\rho\rho}_{ij}\delta_{IJ}D_\rho x^I_{ij}D_\rho x^J_{ji} +c_{ij}^{yy} y_{ij} y_{ji}\right) \\
&(\Delta_2^{(2)})_{ii} = -\gamma^{S^3}\rho^6 \left( -c_{ij}^{yy} y_{ij} y_{ji}+t^{xx}_{ij} \eta^{\mu\alpha}\eta^{\nu\beta} (f_{\mu\nu})_{ij}(f_{\alpha\beta})_{ji}+t^{\rho\rho}_{ij}\eta^{\mu\nu}(f_{\rho\mu})_{ij}(f_{\rho\nu})_{ji}\right)  \ ,
\end{align}
where $\gamma^{S^3}$ is the determinant of the unit $S^3$ metric, $(f_{ab})_{ij}=\partial_a (a_b)_{ij}-\partial_b (a_a)_{ij}$, and
\begin{equation}
    D_a (x^1)_{ij}=D_a y_{ij}=\partial_a y_{ij}-i(m_i-m_j)(a_a)_{ij}\ , \quad D_a (x^2)_{ij}=D_a z_{ij}=\partial_a z_{ij} \ .
\end{equation}
The coefficients appearing in the expansion of the determinant take the values
\begin{align}
c_{ij}^{xx}=& \frac{\lambda}{\pi}\frac{1}{(\rho^2+m_i^2)(\rho^2+m_j^2)} \\
\,c_{ij}^{\rho\rho} =& \frac{\rho^2+m_i^2}{\rho^2+m_j^2} \\
\,c_{ij}^{yy} =& \frac{4(m_i+m_j)^2}{(\rho^2+m_i^2)(\rho^2+m_j^2)} \\
t_{ij}^{xx} =& \frac{\lambda}{\pi} \frac{1}{840}\left[\frac{210}{(\rho^2+m_i^2)^2}+\frac{94}{(\rho^2+m_j^2)^2}+\frac{116}{(\rho^2+m_i^2)(\rho^2+m_j^2)}\right. \notag\\ 
&\left. +(m_i^2-m_j^2)\left(\frac{28}{(\rho^2+m_j^2)^3}-\frac{52}{(\rho^2+m_i^2)^3}\right)+(m_i^2-m_j^2)^2\left(\frac{5}{(\rho^2+m_i^2)^4}+\frac{3}{(\rho^2+m_j^2)^4}\right) \right] \\
t_{ij}^{\rho \rho} =& 1+\frac{1}{140}\frac{(m_i^2-m_j^2)^2}{(\rho^2+m_i^2)^3(\rho^2+m_j^2)^3}\sum_{k=0}^4 a_k(\rho^2+m_i^2)^k(\rho^2+m_j^2)^{4-k} \ , \notag\\ &\text{where}\quad a_0=a_4=1,\ a_1=a_3=10,\ a_2=48.
\end{align}

\subsection{The action to quadratic order in the fluctuations}

We now have the ingredients to expand the action to quadratic order. In order to do this, we provide some simplified formulas for the expansion of the square root of the determinants appearing in the DBI action. First, we have
\begin{equation}
    \left( \det Q^I_{\ J}\right)^{1/2}\sim \mathbb{1}_{\Nf}+\frac{1}{2}Q^{(2)}\ .
\end{equation}
For the determinant along the worldvolume directions 
\begin{equation}
\begin{split}
  \left(-\det K_{ab}\right)^{1/2}&\sim   \left(-\Delta_0-\Delta_1^{(1)}-\Delta_1^{(2)}-\Delta_2^{(2)}\right)^{1/2}=  \left[(-\Delta_0)^{1/2}\left(\mathbb{1}_{\Nf}-\widetilde \Delta_1^{(1)}-\widetilde\Delta_1^{(2)}-\widetilde\Delta_2^{(2)}\right) (-\Delta_0)^{1/2} \right]^{1/2} \\
  &\sim (-\Delta_0)^{1/4}\left[\mathbb{1}_{\Nf}-\frac{1}{2}\left(\widetilde \Delta_1^{(1)}+\widetilde\Delta_1^{(2)}+\widetilde\Delta_2^{(2)}\right) -\frac{1}{8}\widetilde\Delta_1^{(1)}\widetilde\Delta_1^{(1)}\right](-\Delta_0)^{1/4}\\
  &=(-\Delta_0)^{1/2}-(-\Delta_0)^{1/4}\left[\frac{1}{2}\left(\widetilde \Delta_1^{(1)}+\widetilde\Delta_1^{(2)}+\widetilde\Delta_2^{(2)}\right) +\frac{1}{8}\widetilde\Delta_1^{(1)}\widetilde\Delta_1^{(1)}\right](-\Delta_0)^{1/4} \ ,
  \end{split}
\end{equation}
where we have defined
\begin{equation}
    \widetilde \Delta^{(n)}_k=(-\Delta_0)^{-1/2}\Delta^{(n)}_k (-\Delta_0)^{-1/2} \ .
\end{equation}
At the level of components: $(\widetilde \Delta^{(2)}_{1,2})_{ii}=((-\Delta_0)^{-1/2})_{ii}(\Delta^{(2)}_{1,2})_{ii} ((-\Delta_0)^{-1/2})_{ii} = \rho^{-6}(\gamma^{S^3})^{-1}(\Delta^{(2)}_{1,2})_{ii}$. 

Since $\Delta_0\propto \mathbb{1}_{\Nf}$, and the action is quadratic in the fluctuations, the symmetric trace reduces to the ordinary trace. Therefore, using the cyclic property of the trace, we can write
\begin{equation}
\begin{split}
    S_\text{D7}\sim -\frac{1}{(2\pi)^3\gs}\int \dd ^8\sigma \,\text{Tr}_{\Nf}\left[(-\Delta_0)^{1/2}\left(\mathbb{1}_{\Nf}+\frac{1}{2}Q^{(2)}-\frac{1}{2}\left(\widetilde \Delta_1^{(1)}+\widetilde\Delta_1^{(2)}+\widetilde\Delta_2^{(2)}\right) -\frac{1}{8}\widetilde\Delta_1^{(1)}\widetilde\Delta_1^{(1)}\right)\right] \ ,
\end{split}
\end{equation}
where the coordinates and the fields are rescaled as in \eqref{eq:coordtransf}, and the $\lambda$-dependence is encoded in the coefficients.

Taking the trace, and using that $\Delta_1^{(1)}=0$,\footnote{Note that $\Delta_1^{(1)}=0$ because we turned off diagonal fluctuations. However, even if they were non-zero, there would not be any mixing with the off-diagonal fluctuations.} gives
\begin{equation}
    S_\text{D7}=-\frac{1}{(2\pi)^3\gs} \int \dd^4x\int \dd \Omega_3\int \dd \rho\, \rho^3\left[\Nf-\frac{1}{2}\sum_{i}\left((\widetilde\Delta_1^{(2)})_{ii}+(\widetilde\Delta_2^{(2)})_{ii}-q^{(2)}_{ii}\right) \right] \ .
\end{equation}
Since we are considering fluctuations that do not depend on the $S^3$ directions, we can integrate over the sphere ($V(S^3)=2\pi^2$), in such a way that the Lagrangian density to quadratic order is 
\begin{equation}\label{eq:Lquadratic}
\begin{split}
    {\cal L}^{(2)}_\text{D7}=&-\frac{1}{8\pi \gs}\rho^3\sum_{i\neq j}\left[c^{xx}_{ij}\eta^{\mu\nu}\delta_{IJ}D_\mu x^I_{ij}D_\nu x^J_{ji}+c^{\rho\rho}_{ij}\delta_{IJ}D_\rho x^I_{ij}D_\rho x^J_{ji}+c_{ij}^{zz} z_{ij}z_{ji}\right.\\
    &\left.+t^{xx}_{ij} \eta^{\mu\alpha}\eta^{\nu\beta} (f_{\mu\nu})_{ij}(f_{\alpha\beta})_{ji}+t^{\rho\rho}_{ij}\eta^{\mu\nu}(f_{\rho\mu})_{ij}(f_{\rho\nu})_{ji}\right] \ .
\end{split}
\end{equation}
The full action is symmetric under the exchange of flavor indices. Note that some of the coefficients, but not all, already have this symmetry before we sum over all components.

\section{Meson spectra}\label{sec:spectra}

The meson spectrum is obtained by solving the equations of motion derived from the D7-brane action with suitable boundary conditions. The spectrum of the lowest lying heavy-heavy and light-light mesons is known analytically \cite{Kruczenski:2003be}
\begin{equation}\label{eq:mesonmassdiagonal}
    \mathbf{M}_\text{meson} = \sqrt{2\pi}\,\ls\dfrac{2m}{L^2}\sqrt{(n+1)(n+2)}  =2\sqrt{\dfrac{\pi}{\lambda}}\,\mathbf{m_q}\sqrt{(n+1)(n+2)}\ ,
\end{equation}
where we considered the case of zero angular momentum along the three-sphere. We recall that $m$ is the dimensionless position of a single flavor D7-brane, while $\mathbf{m_q}$ is the associated quark mass. We can observe that the meson mass scale is suppressed by a factor of the 't Hooft coupling compared with the quark mass $\mathbf{M}_\text{meson}\sim 1/L\sim \mathbf{m_q}/\sqrt{\lambda}\ll \mathbf{m_q}$. Highly excited mesons with masses of the order of the quark mass are described semiclassically by open Nambu--Goto strings with endpoints at the same flavor D-brane and length of the order of the $AdS$ radius. One can then understand the lower-lying modes as open strings of lengths of order $\sim \ls/\lambda^{1/4}$.\footnote{The energy of a string of length $\ell$ is 
$$
E_{\rm string}=\frac{\ell}{2\pi \ls^2}=\sqrt{\frac{\lambda}{\pi}}\frac{\ell}{L}\frac{1}{L}\,.
$$
The mass of the quark is that of a string of length $\ell\sim L$, so that $\mathbf{m_q}\sim \sqrt{\lambda}/L$. On the other hand, for $\ell\sim \ls/\lambda^{1/4}$ one finds $E_{\rm string} \sim 1/L\sim \mathbf{m_q}/\sqrt{\lambda}$, which is the scale of the meson masses.
}

For heavy-light mesons, the associated open strings will have endpoints at different branes, so if these are separated a distance of the order of the $AdS$ radius, the meson mass scale is expected to be of the same order as the quark mass scale. In order to have a meson spectrum with masses comparable to the lower lying states of the heavy-heavy and light-light mesons, in the large-$\lambda$ limit, the flavor branes should be separated by a distance that is only of the order of $\ls/\lambda^{1/4}$, much smaller than their distance to the stack of color D3-branes. We will study the spectrum in this regime and its behavior as the separation between the flavor branes becomes larger.

\subsection{Equations of motion}

First, we proceed to derive the equations of motion for the fluctuations $\chi_{ij} = \{y_{ij},z_{ij},(a_\mu)_{ij},(a_\rho)_{ij}\}$. To do so, we expand them in Fourier modes
\begin{align}
    \chi_{ij}(x^\mu,\rho)=\int\dfrac{\dd^4k}{(2\pi)^4}e^{ik\cdot x}\chi_{ij}(k_\mu,\rho) \ ,
\end{align}
where $k\cdot x = k_\mu x^\mu$, contracted with the flat metric $\eta_{\mu\nu}$. The Minkowski components of the gauge field in Fourier space can be decomposed into the transverse and longitudinal components
\begin{align}
    A_\mu(\rho,k) = \left(P_{\mu\nu}^\perp+ P_{\mu\nu}^{||}\right)A^\nu(\rho,k) \ ,
\end{align}
where $P_{\mu\nu}^\perp$ and $P_{\mu\nu}^{||}$ are respectively the transverse and longitudinal projectors
\begin{align}
    &P_{\mu\nu}^\perp = \eta_{\mu\nu}-\dfrac{k_\mu k_\nu}{k^2} \ , & &P_{\mu\nu}^{||}=\dfrac{k_\mu k_\nu}{k^2} \ .&
\end{align}
Thus, we can write
\begin{align}\label{eq:amudecomposition}
    A_\mu(\rho, k)  = \sum\limits_{s= 1}^3\zeta_\mu^{(s)}\Phi^{(s)} + k_\mu\Psi \ ,
\end{align}
where $\Phi^{(s)}$ are the three transverse components and $\Psi$ is the longitudinal component of the gauge field. The vectors $\zeta_\mu^{(s)}$ are  the polarization vectors and they satisfy
\begin{align}
 &\zeta^{(s)}\cdot k = 0& &\text{and}& &\zeta^{(s)}\cdot \zeta^{(s')}=\delta_{ss'}\ .&
\end{align}
The Euler--Lagrange equation for the generic fluctuation $\chi_{ij}$ reads
\begin{align}
    &\dfrac{\partial\mathcal{L}_\text{D7}^{(2)}}{\partial\chi_{ji}} =  \partial_\gamma\dfrac{\partial\mathcal{L}_\text{D7}^{(2)}}{\partial\partial_\gamma\chi_{ji}} + \partial_\rho\dfrac{\partial\mathcal{L}_\text{D7}^{(2)}}{\partial\partial_\rho\chi_{ji}}\ ,& &\text{with}\quad\quad\chi_{ij} = \{y_{ij},z_{ij},(a_\mu)_{ij},(a_\rho)_{ij}\}\ .& 
\end{align}
The explicit form of the equations for each component is the following:
\begin{itemize}
    \item Equation of motion for $z_{ij}$
    \begin{align}
        \partial_\rho\left(\rho^3\left(c^{\rho\rho}_{ij}+c^{\rho\rho}_{ji}\right)\partial_\rho z_{ij}\right)-2k^2\rho^3c_{ij}^{xx}z_{ij} =2\rho^3c^{zz}_{ij}z_{ij} \ ,
    \end{align}
    where we used that $c^{xx}_{ij} = c^{xx}_{ji}$, and $c^{zz}_{ij} = c^{zz}_{ji}$.
    \item Equation of motion for $y_{ij}$
    \begin{align}
        \partial_\rho\left[\rho^3\left(c^{\rho\rho}_{ij}+c^{\rho\rho}_{ji}\right)D_\rho y_{ij}\right] +2i\rho^3 c^{xx}_{ij}k^\mu D_\mu y_{ij} = 0\ ,
    \end{align}
     where we used again that $c^{xx}_{ij} = c^{xx}_{ji}$. We can expand the covariant derivatives of $y_{ij}$ and use the decomposition \eqref{eq:amudecomposition} of $a_\mu$ to get
     \begin{align}
         \partial_\rho\left[\rho^3\left(c^{\rho\rho}_{ij}+c^{\rho\rho}_{ji}\right)\left(\partial_\rho y_{ij}-i(m_i-m_j)(a_\rho)_{ij}\right)\right] -2\rho^3 c^{xx}_{ij}k^2(y_{ij}-i(m_i-m_j)\psi_{ij}) = 0 \ .
     \end{align}
    \item Equation of motion for $(a_\mu)_{ij}$
    \begin{align}
        \eta^{\mu\nu}\left\{\partial_\rho\left[\rho^3t^{\rho\rho}_{ij}(f_{\rho\nu})_{ij}\right]-i\rho^3\left(t^{xx}_{ij}+t^{xx}_{ji}\right)\eta^{\gamma\tau}k_\gamma(f_{\nu\tau})_{ij}-i\rho^3(m_i-m_j)c^{xx}_{ij}D_\nu y_{ij}\right\} = 0 \ .
    \end{align}
    where we used that $t^{\rho\rho}_{ij} = t^{\rho\rho}_{ji}$, and $c^{xx}_{ij} = c^{xx}_{ji}$. We can project this equation of motion along the direction transverse to the momentum $k_\mu$ or along the momentum itself. In the first case, we get the equation of motion for the transverse component of the gauge field $\phi^{(s)}_{ij}$:
    \begin{align}
\partial_\rho\left[\rho^3t^{\rho\rho}_{ij}\partial_\rho\phi^{(s)}_{ij}\right]-k^2\rho^3\left(t^{xx}_{ij}+t^{xx}_{ji}\right)\phi^{(s)}_{ij}-\rho^3(m_i-m_j)^2c^{xx}_{ij}\phi^{(s)}_{ij} = 0\ .
    \end{align}
    The second choice leads to the equation for the longitudinal mode of the gauge field $\psi_{ij}$
    \begin{align}
\partial_\rho\left[\rho^3t^{\rho\rho}_{ij}\left(\partial_\rho\psi_{ij}-i(a_\rho)_{ij}\right)\right]+\rho^3(m_i-m_j)c^{xx}_{ij}\left(y_{ij}-(m_i-m_j)\psi_{ij}\right) = 0 \ .
    \end{align}
    \item Equation of motion for $(a_\rho)_{ij}$
    \begin{align}
        2t^{\rho\rho}_{ij}k^\mu(f_{\rho\mu})_{ij}+(m_i-m_j)\left(c^{\rho\rho}_{ij}+c^{\rho\rho}_{ji}\right)D_\rho y_{ij} = 0\ ,
    \end{align}
    where we used $t^{\rho\rho}_{ij} = t^{\rho\rho}_{ji}$. Expanding the covariant derivatives and using \eqref{eq:amudecomposition}, we rewrite the equation as
    \begin{align}
       k^2t^{\rho\rho}_{ij}\left(\partial_\rho\psi_{ij}-i(a_\rho)_{ij}\right)+ (m_i-m_j)\left(c^{\rho\rho}_{ij}+c^{\rho\rho}_{ji}\right)\left(\partial_\rho y_{ij}-i(a_\rho)_{ij}\right) = 0\ .
    \end{align}
\end{itemize}
This completes the set of equations for the flavor brane fluctuations. The scalar $y_{ij}$ mixes with the longitudinal components of the gauge field, leading to Higgsing of these longitudinal modes, as previously observed in \cite{Erdmenger:2007vj}. Since this sector decouples from the transverse fluctuations we are interested in, we will not compute its meson spectrum here.

\subsection{Scalar fluctuations}
Let us recall that the equation of motion for $z_{ij}$ in terms of the Lagrangian coefficients reads
\begin{equation}\label{eq:eqzij}
\partial_\rho\left[\rho^3\left(c^{\rho\rho}_{ij}+c^{\rho\rho}_{ji}\right)\partial_\rho z_{ij}\right]-2k^2\rho^3c_{ij}^{xx}z_{ij} =2\rho^3c^{zz}_{ij}z_{ij} \ ,
\end{equation}

If we substitute the explicit expressions of the coefficients, we get the following differential equation in monic form for $z_{ij}$
\begin{align}\nonumber
    &\partial_\rho^2z_{ij} +\bigg[\dfrac{3}{\rho}-\dfrac{2\rho}{\rho^2+m_{i}^2}-\dfrac{2\rho}{\rho^2+m_{j}^2}+\dfrac{4\rho(2\rho^2+m_{j}^2+m^2_{i})}{(\rho^2+m_{i}^2)^2+(\rho^2+m_{j}^2)^2}\bigg]\partial_\rho z_{ij}\\  \label{eq:eqzzz}
    &+\dfrac{\lambda}{2\pi}\bigg[\dfrac{m_{i}-m_{j}}{m_{i}+m_{j}}\left(\dfrac{1}{\rho^2+m_{i}^2}-\dfrac{1}{\rho^2+m_{j}^2}\right)+\dfrac{4M_\text{meson}^2-2(m_{i}-m_{j})^2}{(\rho^2+m_{i}^2)^2+(\rho^2+m_{j}^2)^2}\bigg]z_{ij}=0 \ ,
\end{align}
where we have introduced the dimensionless meson mass $M_\text{meson}$ as
\begin{equation}\label{eq:mesonrescaling}
    M_\text{meson}^2=-k^2 \,.
\end{equation}
The dimensionless meson mass $M_\text{meson}$ is related to the dimensionful meson mass $\mathbf{M}_\text{meson}$ through
\begin{equation}
  M_\text{meson}  = \sqrt{2\pi} \,\ls \,\mathbf{M}_\text{meson}  \,,
\end{equation}
which is inherited by the coordinate redefinition in equation \eqref{eq:coordtransf}.

\subsubsection{Small quark mass difference}
It is useful to introduce the dimensionless mass difference $\Delta m$ and the average dimensionless mass $\overline{m}$ defined as
\begin{align}\label{eq:massdiffavg}
    &\Delta m = m_{i}-m_{j}\ ,& &\overline{m}=\dfrac{m_{i}+m_{j}}{2}\ .&
\end{align}
In the limit of coincident branes $\Delta m\to 0$, the equation reduces to the equation for scalar fluctuations of \cite{Kruczenski:2003be}, in the case of zero angular momentum along the three-sphere. In this case, $m_i = m_j = m$, and the meson spectrum is given by \eqref{eq:mesonmassdiagonal}.

There is another limit at which the equation has an analytic solution, and hence we can give an analytic expression for the spectrum. In the case of opposite masses $\overline m = 0$, the equation of motion reduces to 
\begin{equation}\label{eq:eqzopposite}
    \partial_\rho^2z_{ij}+\dfrac{3}{\rho}\partial_\rho z_{ij}+\dfrac{16\lambda}{\pi}\dfrac{M_\text{meson}^2-\Delta m^2}{(4\rho^2+\Delta m^2)^2}z_{ij}=0 \ ,
\end{equation}
where $m_i =-m_j =m$, so $\Delta m = 2m$. The meson spectrum then reads 
\begin{equation}\label{eq:mvsmqequation}
     \mathbf{M}_\text{meson}=\sqrt{4\mathbf{m_q}^2+\dfrac{\pi}{\lambda}4\mathbf{m_q}^2(n+1)(n+2)} \ .
\end{equation}
In the large-$\lambda$ limit, the meson mass is proportional to the mass difference $\Delta \mathbf{m_q}$, and hence it is of the order of the quark masses and parametrically larger than the masses of heavy-heavy and light-light mesons. The case of opposite masses corresponds to the antipodal D7-brane configuration relative to the D3-branes. The simplification of the differential equation arises from the fact that the mass-squared matrix becomes proportional to the identity in this case.

However, we are going to show that, if the mass difference is very small, $\Delta m\ll 1$, then the heavy-light meson mass lies between the heavy-heavy and light-light meson masses at finite t'Hooft coupling. In order to show this, we use the theory of perturbations on the equation of motion for $z_{ij}$, where $\Delta m$ is the small parameter. Then, we expand the meson mass and the solution in powers of $\Delta m$:
\begin{align}
   &M_\text{meson}  =  M_\text{meson}^{(0)}+
    M_\text{meson}^{(1)}\Delta m  +M_\text{meson}^{(2)}\Delta m^2+\ldots \\
    &z_{ij} =  z_{ij}^{(0)}+ z_{ij}^{(1)}\Delta m+z_{ij}^{(2)}\Delta m^2+\ldots \ .
\end{align}
At zeroth order in this expansion, we can find an analytic solution for the full spectrum, which is
\begin{equation}\label{eq:m0meson}
    \mathbf{M}_\text{meson}^{(0)} = \sqrt{2\pi}\,\ls\dfrac{2\overline{m}}{L^2}\sqrt{(n+1)(n+2)}  = 2\sqrt{\dfrac{\pi}{\lambda}}\, \mathbf{\overline m_q}\sqrt{(n+1)(n+2)}\ .
\end{equation}
Thus, for a small quark mass difference, the heavy-light meson mass lies perfectly in between the heavy-heavy and light-light meson masses. We can compute the corrections for this leading result. Even if it is not possible to obtain an analytic solution for the full meson tower, we can analytically get the correction for fixed $n$. For example, for $n = 0$, we get
\begin{align}
    \dfrac{M_\text{meson}}{\overline{m}} =2\sqrt{\dfrac{2\pi}{\lambda}}+\dfrac{6\pi + 5\lambda}{20\pi\sqrt{2\pi\lambda}}\dfrac{\Delta m^2}{\overline{m}\,^2}+\mathcal{O}\left(\left(\dfrac{\Delta m}{\overline{m}}\right)^3\right)\ .
\end{align}
We can restore the $\ls$ factors to write the dimensionful meson mass in terms of the quark masses $\mathbf{m_q}$:
\begin{equation}\label{eq:scalarsmallm}
    \dfrac{\mathbf{M}_\text{meson}}{\mathbf{\overline{m}_q}} =2\sqrt{\dfrac{2\pi}{\lambda}}+\dfrac{6\pi + 5\lambda}{20\pi\sqrt{2\pi\lambda}}\dfrac{\Delta \mathbf{m_q}^2}{\mathbf{\overline{m}_q}^2}+\mathcal{O}\left(\left(\dfrac{\Delta \mathbf{m_q}}{\mathbf{\overline{m}_q}}\right)^3\right)\ .
\end{equation}
One can show that this results persists at strong coupling if the separation between the branes is small enough. We recall that, in order for the brane position to be of order of the $AdS$ radius $L$, then $m_{i}\sim\lambda^{1/4}$. As a consequence, the leading term in the dimensionless heavy-light meson mass will scale as $\lambda^{-1/4}$. Moreover, in order to have a brane separation of order the string length $\ls/\lambda^{1/4}$ as previously discussed, the dimensionless mass difference $\Delta m$ should scale as $\lambda^{-1/4}$. Therefore, the ratio $r = \Delta m/\overline m\sim \lambda^{-1/2}$. We can also observe that we can rescale the coordinate $\rho$ as well as $\Delta m$ and $M_\text{meson}$ in units of the dimensionless average quark mass. Then, we can employ the following expansion in powers of $\lambda$:
\begin{align}
    &r =\frac{\widetilde{r}}{\sqrt{\lambda}}\\
    &\dfrac{M_\text{meson}}{\overline m}   =  \dfrac{1}{\sqrt{\lambda}} a^{(0)}(\tilde r)+
    \dfrac{1}{\lambda} a^{(1)}(\tilde r)  +\dfrac{1}{\lambda^{3/2}} a^{(2)}(\tilde r)+\ldots \nonumber\\  
    &z_{ij}  =  z_{ij}^{(0)}+ \dfrac{1}{\sqrt{\lambda}}z_{ij}^{(1)}+\dfrac{1}{\lambda}z_{ij}^{(2)}+\ldots \ .\nonumber
\end{align}
In the large-$\lambda$ limit, the equation of motion at leading order can be solved analytically for the full meson tower, and the spectrum is
\begin{align}\label{eq:mesonspectrumlambda}
    \dfrac{\mathbf{M}_\text{meson}}{\mathbf{\overline m_q}}  =2\sqrt{\dfrac{\pi}{\lambda}}\sqrt{\dfrac{\widetilde{r}\,^2}{4\pi}+(n+1)(n+2)}\,\ , \quad\quad \lambda\to\infty \ .
\end{align}

This formula shows that for small quark mass differences, the mass of heavy-light mesons lies between the heavy-heavy and light-light meson masses, and in particular $\mathbf{M}_\text{meson}\sim \mathbf{\overline m_q}/\sqrt{\lambda}$.

\subsubsection{Numerical solutions}\label{subsubNumerics}

In this section we will solve numerically the equation of motion and compute the exact heavy-light spectrum at fixed values of the 't Hooft coupling. To do so, we use the shooting method, numerically solving the equation of motion from the origin of the spacetime $\rho=0$, and then from the spacetime boundary $\rho = \infty$. 
For the shooting from the origin, we use a cutoff $\rho = 10^{-6}$, and we employ a regular expansion to give an initial value for the solution and its first derivative
\begin{align}\label{eq:scalarIRexpansion}
     z_{ij}|_\text{origin} = \sum\limits_{n=0}^{N_o}a_{2n}\rho^{2n} ,
\end{align}
where $N_o =5$. Fixing $a_0=1$, the first coefficients of the expansion are given in appendix \ref{app:numericsscalar}.

For the shooting from the boundary of the spacetime, we introduce a cutoff at $\rho = 50$. To give initial values for the solution at the cutoff, we use the following series expansion
\begin{align}\label{eq:scalarUVexpansion}
  z_{ij}|_\text{boundary} = \sum\limits_{n=1}^{N_b}b_{2n}\rho^{-2n} \ ,
\end{align}
where $N_b= 6$. Fixing $b_2=1$, the first coefficients of the expansion are given in appendix \ref{app:numericsscalar}.

In this way, we computed the solution first by shooting from the origin $z_{ij}^\text{origin}$ and then from the boundary $z_{ij}^\text{boundary}$. For a given value of the mass parameters $\overline m$ and $\Delta m$, these two solutions smoothly match at an arbitrary value of the holographic coordinate $\rho = \rho_\text{match}$ for only a discrete set of values of the heavy-light meson mass $M_\text{meson}$. This is the set of values that renders the Wronskian of the solutions zero. Therefore, to compute numerically the meson mass, we define the Wroskian as 
\begin{equation}
    \mathcal{W}[\lambda,\overline m,\Delta m, M_\text{meson};\rho] = z^\text{origin}_{ij}(\rho)\partial_\rho z_{ij}^\text{boundary}(\rho)-z^\text{boundary}_{ij}(\rho)\partial_\rho z_{ij}^\text{origin}(\rho)\ ,
\end{equation}
and we use the \textbf{FindRoot} routine of \textit{Mathematica} to find the zeros of the Wronskian $\mathcal{W}$. For the numerical computation, we chose $\rho_\text{match} = 10$ and we checked that the result, $M_\text{meson}$, is independent of this value.

\begin{figure}
\center
\begin{tabular}{cc}
Scalar fluctuations &  Transverse vector fluctuations \\
\includegraphics[width = 0.45\textwidth]{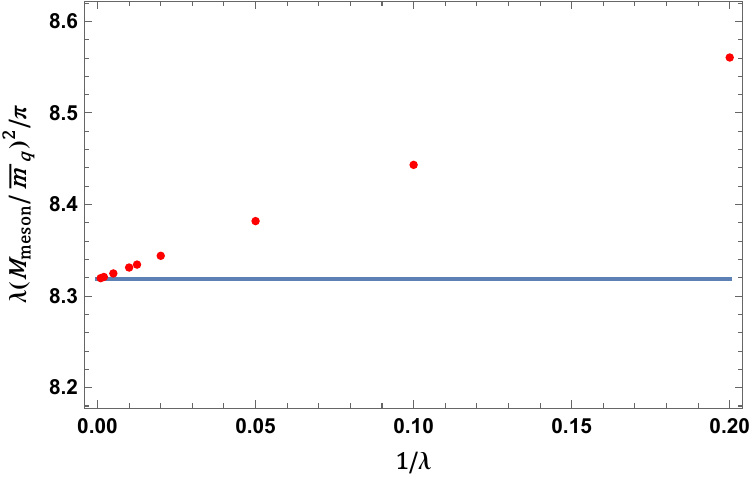} &
\includegraphics[width = 0.45\textwidth]{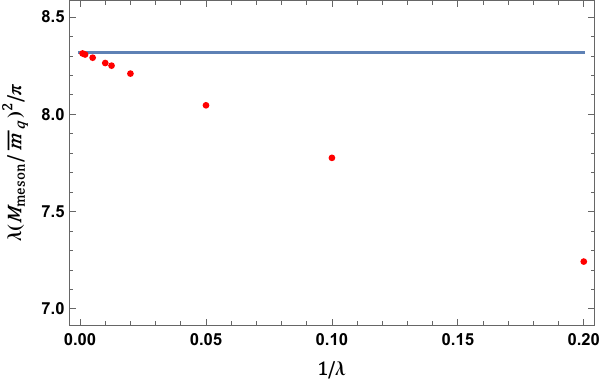}
\end{tabular}
\caption{The plots show the scalar (left panel) and transverse vector (right panel) meson spectrum squared (times $\lambda/\pi$), normalized by the average quark mass $\mathbf{\overline m_q}$, as a function of $1/\lambda$. The solid line corresponds to the analytic result \eqref{eq:mesonspectrumlambda} for $\widetilde{r} = 1$, and $n = 0$ in both plots. The red dots are the numerical results for the meson spectrum.}
\label{fig:scalarvectorspectrumlambda}
\end{figure}

We present the numerical results. First, we analyze the meson spectrum in the large-$\lambda$ limit and compare it to the analytic formula derived in \eqref{eq:mesonspectrumlambda}. In order to do so, we fix the ratio $r = \Delta m/\overline m=\lambda^{-1/2}$, meaning $\tilde r= 1$. Then, we compute the lowest energetic meson mass for several values of $\lambda = \{5,10,20,50,80,100,200,500,1000\}$. The red dots shown in figure~\ref{fig:scalarvectorspectrumlambda} correspond to such meson mass squared normalized by the average quark mass $\mathbf{\overline m_q}$ (and multiplied by $\lambda/\pi$). We compare them with the analytic result (the blue solid line in figure~\ref{fig:scalarvectorspectrumlambda}) that we found in the large-$\lambda$ limit \eqref{eq:mesonspectrumlambda} for $\widetilde{r} = 1$ and $n=0$. As $\lambda$ is increased ($1/\lambda$ decreases in the plot), the numerical solution tends to the approximate analytic solution \eqref{eq:mesonspectrumlambda}. 

Let us adopt the following notation for the D7-brane positions
\begin{align}
    &m_1\equiv m_\text{heavy}\ ,& &m_2\equiv m_\text{light}\ ,&
\end{align}
introduced in the schematic representation of the branes setup in figure \ref{fig:setup}. We will then refer to the physical quark masses as
\begin{align}
    &\mathbf{m_{q,\text{heavy}}}\equiv (2\pi \ls^2)^{-1/2}\,m_\text{heavy} \ ,& &\mathbf{m_{q,\text{light}}}\equiv (2\pi \ls^2)^{-1/2}\, m_\text{light}\ .&
\end{align}

\begin{figure}
\center
\begin{tabular}{cc}
$\lambda=2^4$ &  $\lambda=3^4$ \\
\includegraphics[width = 0.5\textwidth]{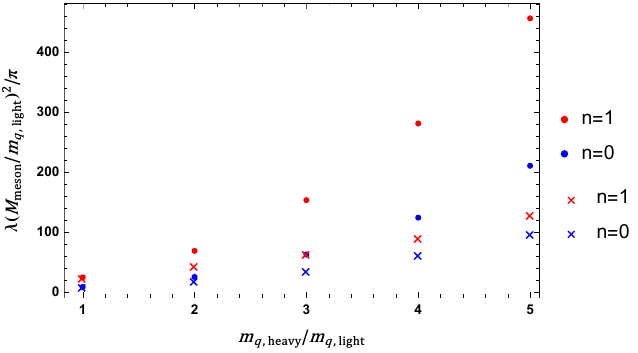} &
\includegraphics[width = 0.5\textwidth]{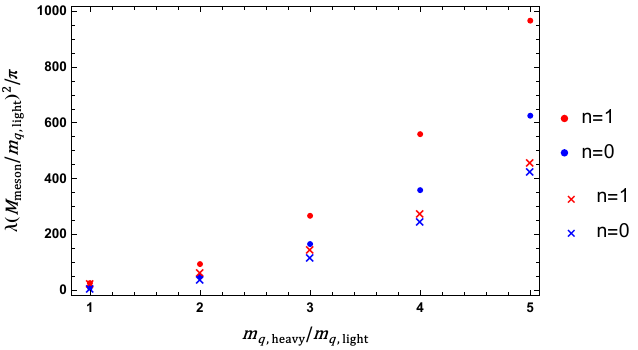}
\end{tabular}
\caption{The plot shows the $n=0$ (blue) and $n=1$ (red) scalar meson mass squared (times $\lambda/(\mathbf{m}^2_\mathbf{{q,\text{light}}}\pi)$) as a function of the mass of the heavy quark $\mathbf{m_{q,\text{heavy}}}$ (rescaled by $\mathbf{m_{q,\text{light}}}$), for fixed $\lambda=2^4$ (left panel) and $\lambda=3^4$ (right panel). The blue and red dots correspond respectively to the $n=0$ and $n=1$ heavy-light meson mass computed from \eqref{eq:eqzzz}, while the blue and red crosses correspond respectively to the $n=0$ and $n=1$ masses obtained in \cite{Erdmenger:2007vj}.}
\label{fig:scalar}
\end{figure}

Next, we will fix $\lambda$ and compute the meson spectrum as a function of the ratio $ m_\text{heavy}/m_\text{light}$, which is equal to the ratio of the heavy and light quark masses $ \mathbf{m_{q,\text{heavy}}}/\mathbf{m_{q,\text{light}}}$, with restored units. We do this because one can see from the equation of motion of $z_{ij}$ \eqref{eq:eqzzz} that we can rescale $\rho$, $m_\text{heavy}$, and $M_\text{meson}$ in units of $m_\text{light}$. In figure~\ref{fig:scalar}, the blue dots represent the $n=0$ heavy-light meson mass $\mathbf{M}_\text{meson}/\mathbf{m_{q,\text{light}}}$ (or just $M_\text{meson}$ for fixed $m_{\text{light}}=1$) as a function of the heavy quark mass (in units of the light quark mass) for fixed $\lambda=2^4$ (left panel) and $\lambda=3^4$ (right panel). We want to mention that for sufficiently low $\lambda$, the heavy-light meson mass is always between the heavy-heavy and light-light masses. However, if we increase $\lambda$, then the heavy-light meson mass becomes the largest. Moreover, even for small values of $\lambda$, increasing the heavy-heavy meson mass raises the mass of the heavy-light mesons in such a way that they eventually become the heaviest modes. The red dots in figure \ref{fig:scalar} are the corresponding results for the first excited $n=1$ heavy-light meson mass.

In \cite{Erdmenger:2006bg,Erdmenger:2007vj}, the meson mass of heavy-light mesons was studied first by investigating the Nambu-Goto action of strings stretching between the flavor D7-branes and then exploring the non-Abelian DBI action for the flavor branes. As we discussed in the introduction, our careful analysis of the DBI action introduces some differences with respect to those older computations. In particular, the equation of motion for the fluctuation $z_{ij}$ in \cite{Erdmenger:2007vj} reads
\begin{equation}\label{eq:eqzijErd}
    \partial_\rho^2z_{ij}+\dfrac{3}{\rho}\partial_\rho z_{ij}+\dfrac{\lambda\left(M_\text{meson}^2-(m_i-m_j)^2\right)}{2\pi}\left(\dfrac{1}{(\rho^2+m_i^2)^2}+\dfrac{1}{(\rho^2+m_j^2)^2}\right)z_{ij}=0 \ ,
\end{equation}
which is clearly different from the equation of motion for $z_{ij}$ that we derived (see equation \eqref{eq:eqzzz}). We stress that for the present equation, the main source of discrepancy comes from our choice of making the induced metric Hermitian. 

Such an equation can be solved numerically using the same procedure we used for \eqref{eq:eqzzz}. Even though the qualitative results for the meson spectrum are the same in the two cases, meaning that the heavy-light meson mass is larger than the heavy-heavy and light-light meson masses for large $\lambda$, quantitatively, the meson spectrum is different. 
In figure~\ref{fig:scalar}, the blue and red crosses correspond respectively to the results for the meson spectrum of \cite{Erdmenger:2007vj} for the $n=0$ and the $n=1$ modes. The plot shows that the meson mass we computed is bigger than the one computed in \cite{Erdmenger:2007vj} for both the $n=0,1$ modes, and for $\lambda=2^4,3^4$, but the upshot holds for a generic $\lambda$. 

\subsection{Transverse vector fluctuations}

In this section, we compute the meson mass associated with the vector sector of the fluctuations. We start by recalling the equation of motion for the transverse component of the gauge field 
\begin{align}\label{eq:eqphiij}
\partial_\rho\left[\rho^3t^{\rho\rho}_{ij}\partial_\rho\phi^{(s)}_{ij}\right]+M_\text{meson}^2\rho^3\left(t^{xx}_{ij}+t^{xx}_{ji}\right)\phi^{(s)}_{ij}-\rho^3(m_i-m_j)^2c^{xx}_{ij}\phi^{(s)}_{ij} = 0\ ,
    \end{align}
where we introduced the dimensionless meson mass $M_\text{meson}^2 = -k^2$. 

We first observe that the equation of motion for $\phi_{ij}$ is not the same as the equation of motion for $z_{ij}$, unlike in \cite{Erdmenger:2007vj}. In that case, the scalar and vector heavy-light meson masses were degenerate since the equations of motion coincided. This is not the case in the present work, and therefore, there will not be a degeneracy of the scalar and vector heavy-light meson masses. We also observe that the differences of the equation of motion for $\phi_{ij}$ with respect to the one derived in \cite{Erdmenger:2007vj} arise both from the Hermitian definition of the matrices and, more crucially, from how we dealt with the determinant.

\subsubsection{Small quark mass difference}

We introduce the mass difference $\Delta m$, and the average mass $\overline m$, as we did for the scalar sector in \eqref{eq:massdiffavg}. We can take the limit of coincident branes $\Delta m\to 0$, and the equation reduces to the equation for vector fluctuations of \cite{Kruczenski:2003be}, in the case of zero angular momentum along the three-sphere. This equation is the same as the equation of motion for scalar fluctuations in the coincident limit. Therefore, the meson spectrum is given by \eqref{eq:mesonmassdiagonal}, with $m_i = m_j = m$.

As for the scalar equation of motion, also the equation of motion for $\phi_{ij}$ has an analytic solution in the case of opposite masses $\overline m = 0$. The equation of motion reduces to 
\begin{equation}
    \partial_\rho^2\phi_{ij}+\dfrac{3}{\rho}\partial_\rho \phi_{ij} +\dfrac{16\lambda}{\pi}\dfrac{M_\text{meson}^2-\Delta m^2}{(4\rho^2+\Delta m^2)^2}\phi_{ij}=0 \ ,
\end{equation}
where $m_i =-m_j =m$, so $\Delta m = 2m$. The equation is the same as the scalar one in the opposite mass limit \eqref{eq:eqzopposite}. Therefore, the meson spectrum is again
\begin{equation}
     \mathbf{M}_\text{meson}=\sqrt{4\mathbf{m_q}^2+\dfrac{\pi}{\lambda}4\mathbf{m_q}^2(n+1)(n+2)} \ .
\end{equation}

Then, we proceed as we did in the previous section for $z_{ij}$ and find the meson spectrum in the limit of small mass difference $\Delta m\ll 1$. We will find that, also for the transverse fluctuation, the heavy-light meson mass lies between the heavy-heavy and light-light meson masses. Let us expand the meson mass and the solution in powers of $\Delta m$:
\begin{align}
    &M_\text{meson}  =M_\text{meson}^{(0)}+
    M_\text{meson}^{(1)}\Delta m  +M_\text{meson}^{(2)}\Delta m^2+\ldots  \\
    &\phi_{ij} = \phi_{ij}^{(0)}+ \phi_{ij}^{(1)}\Delta m+\phi_{ij}^{(2)}\Delta m^2+\ldots  \ .
\end{align}

At zeroth order in this expansion, we have an analytic solution, and the spectrum $\mathbf{M}_\text{meson}^{(0)}$ is given by the same formula as for the scalar sector \eqref{eq:m0meson}. We can also compute the corrections to this leading result in this case. As for the scalar spectrum, we compute analytically the correction for fixed $n$. For example, for $n = 0$
\begin{align}
   \dfrac{M_\text{meson}}{\overline{m}} =2\sqrt{\dfrac{2\pi}{\lambda}}+\dfrac{-26 + 5\lambda}{20\pi\sqrt{2\pi\lambda}}\dfrac{\Delta m^2}{\overline{m}\,^2}+\mathcal{O}\left(\left(\dfrac{\Delta m}{\overline{m}}\right)^3\right)\ .
\end{align}
Restoring the $\ls$ factors, we can write
\begin{equation}\label{eq:vectorsmallmass}
    \dfrac{\mathbf{M}_\text{meson}}{\mathbf{\overline{m}_q}} =2\sqrt{\dfrac{2\pi}{\lambda}}+\dfrac{-26 + 5\lambda}{20\pi\sqrt{2\pi\lambda}}\dfrac{\Delta \mathbf{m_q}^2}{\mathbf{\overline{m}_q}^2}+\mathcal{O}\left(\left(\dfrac{\Delta \mathbf{m_q}}{\mathbf{\overline{m}_q}}\right)^3\right)\ .
\end{equation}

We observe that this result slightly differs from the one obtained in the scalar sector. In particular, the $(\Delta \mathbf{m_q}/\mathbf{\overline m _q})^2$-correction has a negative term which is not proportional to $\lambda$. On the contrary, the same correction for the meson spectrum \eqref{eq:scalarsmallm} has a positive term not proportional to $\lambda$. As a consequence, for $\lambda$ not very large and a small mass separation, the vector mesons are lighter than the scalar mesons. This is in stark contrast with what happens in QCD, where scalar mesons are lighter than their vector partners. However, we know that, in QCD, this happens because scalar mesons are protected by the approximate chiral symmetry, which is absent in $\mathcal{N}=4$ SYM with massless quark degrees of freedom. As we did for the scalar mesons, we can study the equation of motion in the large-$\lambda$ limit. We can rescale the coordinate $\rho$ as well as $\Delta m$ and $M_\text{meson}$ in units of the dimensionless average mass $\overline m$. Then, we can employ the following expansion in powers of $\lambda$:
\begin{align}
    &r =\frac{\widetilde{r}}{\sqrt{\lambda}}\\
    &\dfrac{M_\text{meson}}{\overline m}   =  \dfrac{1}{\sqrt{\lambda}} a^{(0)}(\tilde r)+
    \dfrac{1}{\lambda} a^{(1)}(\tilde r)  +\dfrac{1}{\lambda^{3/2}} a^{(2)}(\tilde r)+\ldots \nonumber\\  
    &\phi_{ij}  =  \phi_{ij}^{(0)}+ \dfrac{1}{\sqrt{\lambda}}\phi_{ij}^{(1)}+\dfrac{1}{\lambda}\phi_{ij}^{(2)}+\ldots \ .\nonumber
\end{align}

In the large-$\lambda$ limit, we can solve the equation of motion analytically and compute the spectrum. This coincides with the scalar meson mass \eqref{eq:mesonspectrumlambda}, in this limit.

\subsubsection{Numerical solutions}

The numerical routine for finding the solutions to the equation of motion for $\phi_{ij}$ is the same as the one discussed in section \ref{subsubNumerics}. Here, we briefly adapt it to the vector case. 

We use $\rho=10^{-6}$ and $\rho = 50$ as cutoffs, respectively, for the origin and the boundary of the spacetime. For the shooting from the origin and the boundary, we employ the following series expansions for $\phi_{ij}$
\begin{align}\label{eq:vectorexpansions}
   &\phi_{ij}|_\text{origin}=\sum_{n=0}^{N_o}c_{2n}\rho^{2n} \ , &   &\phi_{ij}|_\text{boundary}=\sum_{n=1}^{N_b}d_{2n}\rho^{-2n} \ ,&
\end{align}
where $N_o=5$, and $N_b=6$. The first coefficients of the expansions, provided that $c_0=d_2=1$, are given in appendix \ref{app:numericsvector}.

With these as boundary conditions, we compute the solution shooting from the origin and from the boundary, $\phi_{ij}^{\text{origin}}$, and $\phi^\text{boundary}_{ij}$. Then, we define the Wronskian of the solutions
\begin{equation}
    \mathcal{W}[\lambda,\overline m,\Delta m, M_\text{meson};\rho] = \phi^\text{origin}_{ij}(\rho)\partial_\rho \phi_{ij}^\text{boundary}(\rho)-\phi^\text{boundary}_{ij}(\rho)\partial_\rho \phi_{ij}^\text{origin}(\rho)
    \end{equation}
and we use the \textbf{FindRoot} routine of \textit{Mathematica} to find the zeros of the Wronskian $\mathcal{W}$ at fixed $\overline m$, and $\Delta m$. These zeros happen to be located at the values of $M_\text{meson}$ for which $\phi_{ij}^{\text{origin}}$ and $\phi^\text{boundary}_{ij}$ smoothly match at an arbitrary value of the holographic coordinate. As for the scalar sector, we adopted $\rho_\text{match}=10$, and checked that the result is independent of this choice.

First, we show that, also in this case, the meson mass in the large-$\lambda$ limit tends to \eqref{eq:mesonspectrumlambda}. We fix the ratio $r = \Delta m/\overline m= \lambda^{-1/2}$, meaning $\tilde r = 1$, and we compute the lowest energetic meson mass squared (times $\lambda/\pi$), normalized by the average quark mass $\mathbf{\overline m_q}$. The numerical results are presented as red dots in figure~\ref{fig:scalarvectorspectrumlambda}, while the blue solid line represents the analytic result \eqref{eq:mesonspectrumlambda}. As one can see from the plot, the vector heavy-light meson mass goes to the analytic result as the scalar one, but from below. This is in agreement with the result obtained in \eqref{eq:vectorsmallmass} due to $\lambda^0$-negative correction. This numerical analysis confirms what we have seen by confronting the approximate analytic results \eqref{eq:scalarsmallm}, \eqref{eq:vectorsmallmass}: the vector heavy-light mesons are lighter than the scalar heavy-light mesons. This breaks the mass degeneracy between the vector and scalar mesons observed in \cite{Erdmenger:2007vj}.

\begin{figure}
\center
\includegraphics[width = 0.65\textwidth]{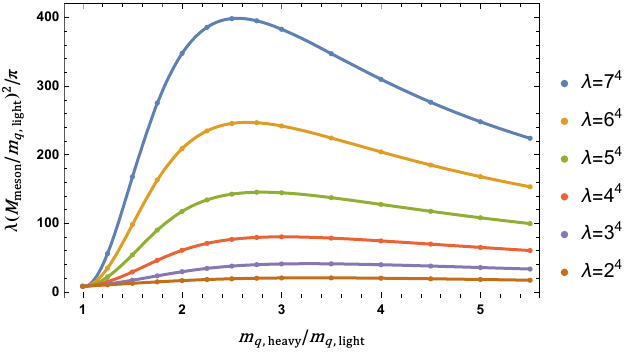}
\caption{The plot shows the mass squared of the lightest transverse vector mode (times $\lambda/( \mathbf{m}^2_\mathbf{{q,\text{light}}}\pi)$) as a function of $\mathbf{m_{q,\text{heavy}}}/\mathbf{m_{q,\text{light}}}$ and several values of the 't Hooft coupling $\lambda = \{2^4,3^4,3^4,4^4,5^4,6^4,7^4\}$. The dots are the numerical results for the meson mass, while the solid curves are interpolations of the data points.}
\label{fig:peakshift}
\end{figure}

Then, as for the scalar equation of motion, we can rescale $\rho$, $m_\text{heavy}$, and $M_\text{meson}$ in units of $m_\text{light}$. We fix $\lambda$ to compute the transverse vector meson spectrum as a function of the heavy quark mass $\mathbf{m_{q,\text{heavy}}}$ (rescaled with $\mathbf{m_{q,\text{light}}}$). In figure \ref{fig:peakshift}, we plot the lowest energy mode $n=0$ of the vector spectrum for $\lambda = \{2^4,3^4,3^4,4^4,5^4,6^4,7^4\}$, as a function of $\mathbf{m_{q,\text{heavy}}}/\mathbf{m_{q,\text{light}}}$. The numerical results are represented by the dots (different colors for different values of $\lambda$), while the solid lines are the corresponding interpolating functions. We computed the maximum of these functions, and we found that the peaks shift to the left as we raise $\lambda$. In particular, the maximum of the curves associated with the values of $\lambda = \{2^4,3^4,3^4,4^4,5^4,6^4,7^4\}$ are respectively $\{3.20,3.10,3.00,2.82,2.66,2.56\}$, which clearly shows the behavior just mentioned. In other words, the lowest energetic vector meson mass starts to decrease with $\mathbf{m_{q,\text{heavy}}}/\mathbf{m_{q,\text{light}}}$ for smaller and smaller $\mathbf{m_{q,\text{heavy}}}/\mathbf{m_{q,\text{light}}}$ as we raise $\lambda$.

\begin{figure}
\center
\begin{tabular}{c}
   $\lambda=2^4$\\
\includegraphics[width = 0.65\textwidth]{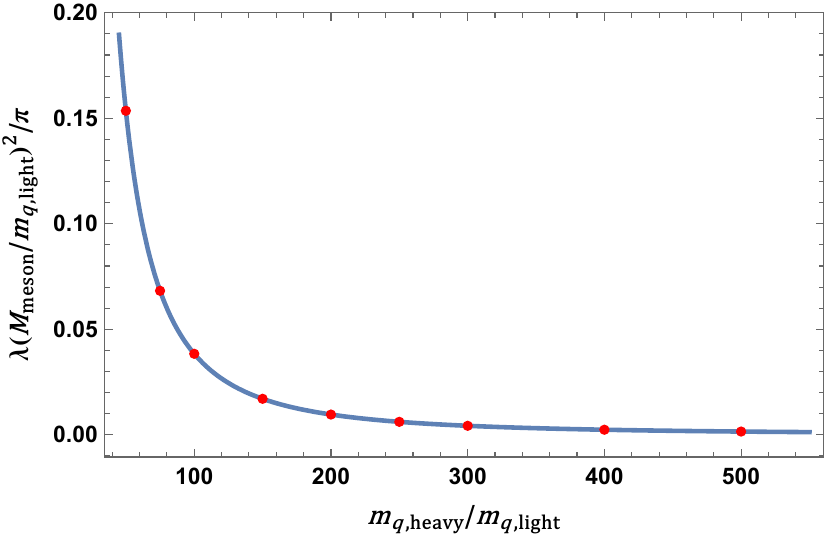}
\end{tabular}
\caption{The plot shows the transverse vector meson spectrum squared (times $\lambda/(\mathbf{m}^2_\mathbf{{q,\text{light}}}\pi)$), for fixed $\lambda=2^4$, as a function of the ratio $\mathbf{m_{q,\text{heavy}}}/\mathbf{m_{q,\text{light}}}$ in the regime where it is very large. The dots correspond to the numerical solutions for $\mathbf{m_{q,\text{heavy}}}/\mathbf{m_{q,\text{light}}}=\{50,75,100,150,200,250,300,400,500\}$, while the blue solid curve is the numerical fit \eqref{eq:massratiofit}.
}
\label{fig:largeheavyquarkmass}
\end{figure}

One can check that the transverse vector meson masses are always (for every value of $\lambda$) lighter than the scalar meson masses. Moreover, thanks to our analysis, we discover that, contrary to expectations, the mass of the first heavy-light meson decreases as the heavy quark mass increases. This is not just a quantitative difference but rather a qualitative difference with respect to the results present in the literature, for which the spectrum is the same as the one presented in figure \ref{fig:scalar}. In appendix \ref{app:numanalysis}, we analyzed in more detail the equation of motion for the transverse fluctuation in the limit of very large heavy quark mass in order to check if the behavior of figure~\ref{fig:peakshift} continues to hold. That is indeed the case, as shown in figure \ref{fig:largeheavyquarkmass}, where the lowest vector meson mass, for $\lambda=2^4$ is shown as a function of $\mathbf{m_{q,\text{heavy}}}/\mathbf{m_{q,\text{light}}}$ in the regime where this ratio is large. The numerical results are represented by red dots. Numerically, we can fit these points with a power-law function (blue solid line). The result reads 
\begin{align}\label{eq:massratiofit}
    \dfrac{\lambda}{\pi}\,\dfrac{\mathbf{M}^2_\text{meson}}{\mathbf{m}^2_\mathbf{{q,\text{light}}}} \approx a^2\,\dfrac{\mathbf{m}^2_\mathbf{{q,\text{light}}}}{\mathbf{m}^2_\mathbf{{q,\text{heavy}}}} \ ,
\end{align}
where $a\approx 288$. See appendix \ref{app:numanalysis} for more details on the derivation of this formula.

\begin{figure}
\center
\begin{tabular}{cc}
$\lambda=2^4$ &  $\lambda=3^4$ \\
\includegraphics[width = 0.5\textwidth]{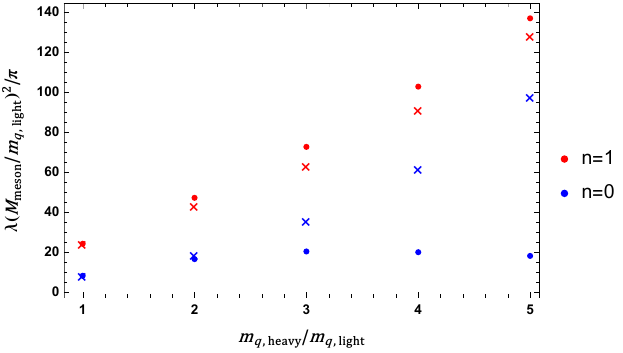} &
\includegraphics[width = 0.5\textwidth]{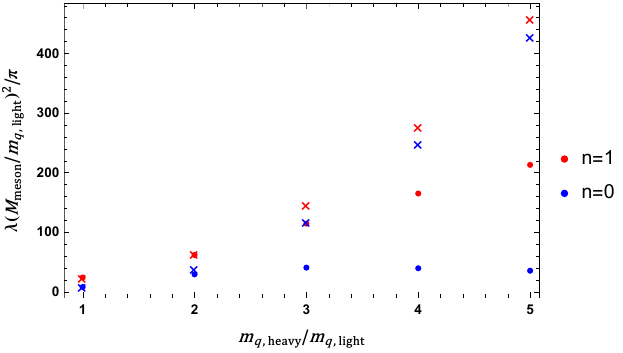}
\end{tabular}
\caption{The plot shows the $n=0$ (blue) and $n=1$ (red) vector meson mass squared (times $\lambda/(\mathbf{m}^2_\mathbf{{q,\text{light}}}\pi)$) as a function of the mass of the heavy quark $\mathbf{m_{q,\text{heavy}}}$ (rescaled by $\mathbf{m_{q,\text{light}}}$), for fixed $\lambda=2^4$ (left panel) and $\lambda=3^4$ (right panel). The blue and red dots correspond respectively to the $n=0$ and $n=1$ heavy-light meson mass computed from \eqref{eq:eqphiij}, while the blue and red crosses correspond respectively to the $n=0$ and $n=1$ masses obtained in \cite{Erdmenger:2007vj}.}
\label{fig:vector}
\end{figure}

We conclude this section by showing the comparison between the $n=0,1$ transverse vector modes' mass and the corresponding masses derived from the equation of motion in \cite{Erdmenger:2007vj}. The latter are the same as the scalar ones since the vector and scalar masses are degenerate. As in the previous section, we fix $\lambda$ and
compute the meson spectrum as a function of the ratio $\mathbf{m_{q,\text{heavy}}}/\mathbf{m_{q,\text{light}}}$. In the figure \ref{fig:vector}, the blue dots represent the $n=0$ heavy-light vector meson mass $\mathbf{M}_\text{meson}/\mathbf{m_{q,\text{light}}}$ as a function of $\mathbf{m_{q,\text{heavy}}}/\mathbf{m_{q,\text{light}}}$ for
fixed $\lambda=2^4$ (left panel) and $\lambda=3^4$ (right panel). The blue and red crosses correspond respectively to the results for the meson spectrum computed from \eqref{eq:eqzijErd}, the equation of motion for the fluctuations as computed in \cite{Erdmenger:2007vj}, for both the $n=0,1$ modes. The zero modes display the peculiar behavior discussed above. The first excited vector meson modes do not show the same behavior. In this case, the meson mass grows as the $\mathbf{m_{q,\text{heavy}}}/\mathbf{m_{q,\text{light}}}$ grows (or $\mathbf{m_{q,\text{heavy}}}$ at fixed $\mathbf{m_{q,\text{light}}}$). We checked this also for larger values of $\mathbf{m_{q,\text{heavy}}}/\mathbf{m_{q,\text{light}}}$ than the one we reported in the plots. Therefore, the appearance of a light meson mode is restricted to the lowest energy mode of the vector spectrum. We also observe that if we raise $\lambda$, our numerical results become lighter than the $n=1$ masses computed from \eqref{eq:eqzijErd}.

\subsection{Comparison with the classical string computation}

In this section, we want to compare our results for the meson spectra, arising from the expansion of the non-Abelian DBI action, with the expected classical spectrum of fluctuations of a fundamental string stretching between the two branes. 

We recall that the D7-branes are placed at $z=0$ and different values of $y$. In order to compute the classical energy of the string, we adopt the static gauge $\tau =t$ and we parametrize the spatial embedding by a parameter $\sigma$. Then, the energy as derived from the Nambu--Goto action reads
\begin{equation}
    \mathbf{E}_\text{string} = \dfrac{1}{2\pi \ls^2}\int \dd \sigma\,\sqrt{-g_{tt}g_{ab}\dfrac{\dd X^a}{\dd \sigma}\dfrac{\dd X^b}{\dd \sigma}} \ ,
\end{equation}
where the metric coefficients refer to the spacetime metric \eqref{eq:metric10d} and $a,b=\rho,\alpha,z$ with $\alpha$ the index which runs along the three-sphere directions.\footnote{We are looking for a static configuration whose endpoints lie at the same point in the $x^\mu$directions, so $x^\mu(\sigma)=\rm{constant}$.} In general, the above energy is just the string tension times the ordinary Euclidean length of the path in the $(\rho,S^3,y,z)$-space:
\begin{equation}
    \mathbf{E}_\text{string} = \dfrac{1}{2\pi \ls^2}\int \dd \sigma\,\sqrt{\dot\rho^2+\rho^2|\dot\theta|^2+\dot y^2+\dot z^2}\ .
\end{equation}
If the two endpoints are at the same $\rho$, same position along $S^3$, and same $z$, the minimal path is trivially a straight line in the $y$ direction. Thus,
\begin{equation}\label{eq:NGenergy}
    \mathbf{E}_\text{string} =\mathbf{m_{q,\text{heavy}}}-\mathbf{m_{q,\text{light}}}\,,
\end{equation}
where we presented the formula in terms of the dimensionful quark masses. 
 
For a fundamental string quantized around this configuration in a semiclassical approximation, we expect to find modes with an energy that equals  \eqref{eq:NGenergy} plus a correction suppressed by the string scale or, equivalently, the 't Hooft coupling. A deviation from this behavior of the meson spectra indicates either that the semiclassical approximation for the string breaks down or that $\alpha'$ corrections to Myers' action become relevant.

In the following, we present the difference between our numerical results for the meson spectra and the classical result coming from the Nambu--Goto energy \eqref{eq:NGenergy}. 

\begin{figure}
\begin{tabular}{cc}
\includegraphics[width = 0.5\textwidth]{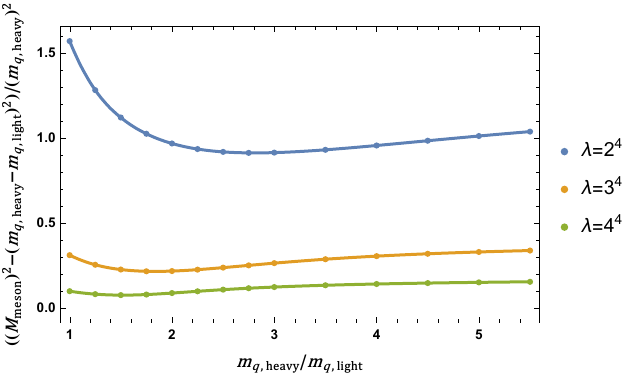} &
\includegraphics[width = 0.5\textwidth]{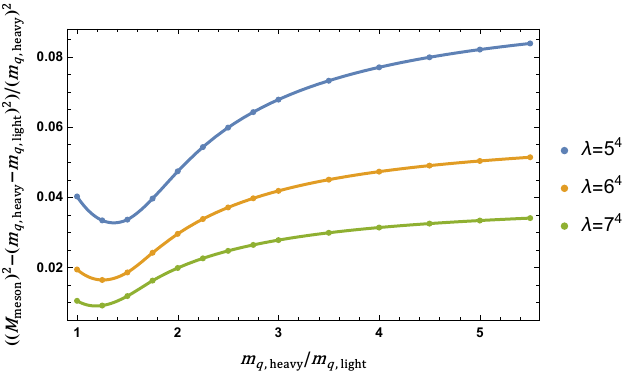}
\end{tabular}
\caption{The plot shows the difference between the $n=0$ scalar meson mass squared and the energy squared of a classical string stretching between the D7-brane (normalized with $\mathbf{m}^{-2}_\mathbf{{q,\text{heavy}}}$) \eqref{eq:difference} as a function of the mass of the ratio $\mathbf{m_{q,\text{heavy}}}/\mathbf{m_{q,\text{light}}}$ at fixed $\lambda$. The left panel shows the results for $\lambda=\{2^4,3^4,4^4\}$, while the right panel for $\lambda=\{5^4,6^4,7^4\}$. In both plots, the dots correspond to the numerical results, while the solid lines are the corresponding interpolating functions.}
\label{fig:scalarsemiclass0}
\end{figure}

We start from the results for the scalar meson spectrum reported in figure \ref{fig:scalarsemiclass0}. Here, we plot the difference
\begin{equation}\label{eq:difference}
    \dfrac{\mathbf{M}_\text{meson}^2-\mathbf{E}_\text{string}^2}{\mathbf{m}^2_\mathbf{{q,\text{heavy}}}} =\dfrac{\mathbf{M}_\text{meson}^2-(\mathbf{m_{q,\text{heavy}}}-\mathbf{m_{q,\text{light}}})^2}{\mathbf{m}^2_\mathbf{{q,\text{heavy}}}} \,,
\end{equation}
as a function of the ratio $\mathbf{m_{q,\text{heavy}}}/\mathbf{m_{q,\text{light}}}$, for several values of $\lambda$. The left and right panel of the figure \ref{fig:scalarsemiclass0} shows the plots respectively for $\lambda=\{2^4,3^4,4^4\}$ and $\lambda=\{5^4,6^4,7^4\}$. The dots correspond to the numerical solutions, while the solid lines are the corresponding interpolating functions. We observe that our numerical results become closer to the classical string energy as we raise $\lambda$, consistently with a semiclassical approximation. However, as we raise the value of the ratio $\mathbf{m_{q,\text{heavy}}}/\mathbf{m_{q,\text{light}}}$, our results deviate slowly from \eqref{eq:NGenergy}.

\begin{figure}
\begin{tabular}{cc}
\includegraphics[width = 0.5\textwidth]{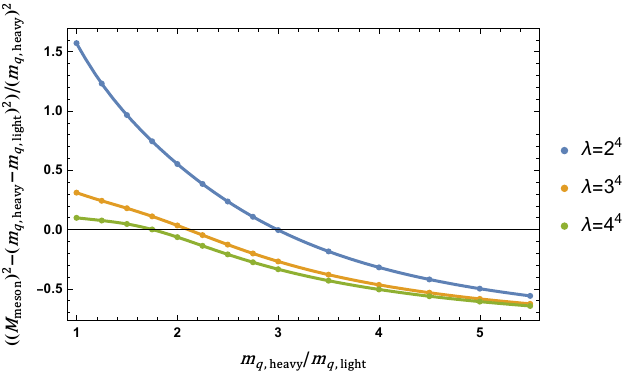} &
\includegraphics[width = 0.5\textwidth]{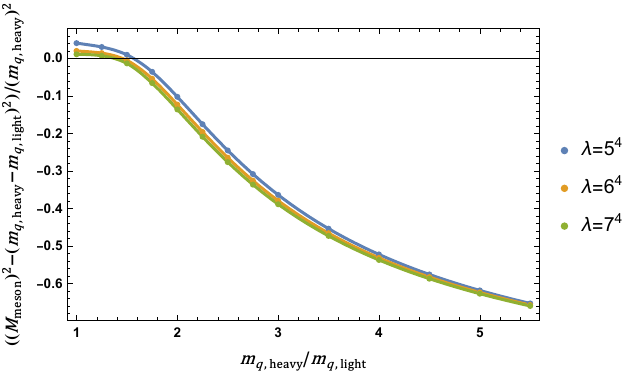}
\end{tabular}
\caption{The plot shows the difference between the $n=0$ transverse vector meson mass squared and the energy squared of a classical string stretching between the D7-brane (normalized with $\mathbf{m}^{-2}_\mathbf{{q,\text{heavy}}}$) \eqref{eq:difference} as a function of the mass of the ratio $\mathbf{m_{q,\text{heavy}}}/\mathbf{m_{q,\text{light}}}$ at fixed $\lambda$. The left panel shows the results for $\lambda=\{2^4,3^4,4^4\}$, while the right panel for $\lambda=\{5^4,6^4,7^4\}$. In both plots, the dots correspond to the numerical results, while the solid lines are the corresponding interpolating functions.}
\label{fig:vectorsemiclass0}
\end{figure}

The situation is different for the zero mode of the transverse meson spectrum, shown in figure \ref{fig:vectorsemiclass0}. In the regime where the meson mass increases, to the left of the peak in figure \ref{fig:peakshift}, the behavior is similar to that of the scalar mode. Afterward, the behavior becomes largely insensitive to the value of $\lambda$ and increasingly deviates from the classical value of the string energy as $\mathbf{m_{q,\text{heavy}}}/\mathbf{m_{q,\text{light}}}$ is increased. To determine whether this is due to a shift of the mass of the string modes relative to flat space or to missing terms in the D-brane action would require further study of the quantized string in this setup.

\begin{figure}
\begin{tabular}{cc}
\includegraphics[width = 0.5\textwidth]{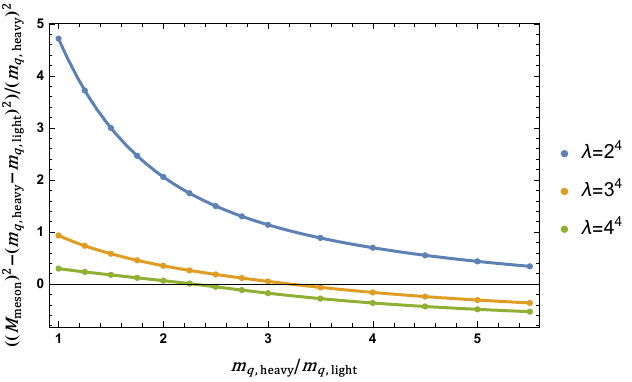} &
\includegraphics[width = 0.5\textwidth]{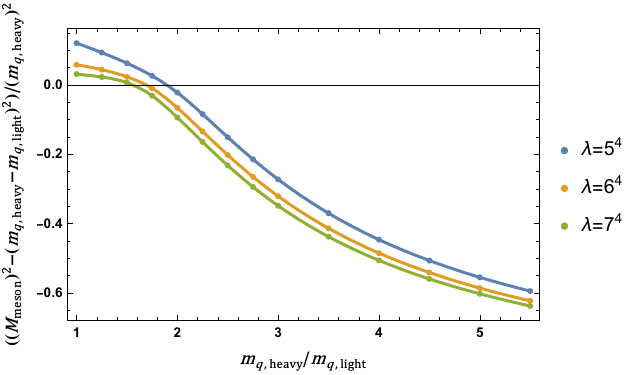}
\end{tabular}
\caption{The plot shows the difference between the $n=1$ transverse vector meson mass squared and the energy squared of a classical string stretching between the D7-brane (normalized with $\mathbf{m}^{-2}_\mathbf{{q,\text{heavy}}}$) \eqref{eq:difference} as a function of the mass of the ratio $\mathbf{m_{q,\text{heavy}}}/\mathbf{m_{q,\text{light}}}$ at fixed $\lambda$. The left panel shows the results for $\lambda=\{2^4,3^4,4^4\}$, while the right panel for $\lambda=\{5^4,6^4,7^4\}$. In both plots, the dots correspond to the numerical results, while the solid lines are the corresponding interpolating functions.}
\label{fig:vectorsemiclass1}
\end{figure}

We can carry out the same comparison for the first excited mode of the transverse vector meson spectrum. Even if in the previous section we saw that such a mode has a mass which grows as we increase the value of the ratio $\mathbf{m_{q,\text{heavy}}}/\mathbf{m_{q,\text{light}}}$, as for the vector $n=0$ mode, it increasingly deviates from the classical value of the string energy as we raise $\lambda$. This can be seen in figure \ref{fig:vectorsemiclass1}, which shows the same behavior as in the corresponding figure \ref{fig:vectorsemiclass0} for the vector mode $n=0$. 

\section{Discussion}\label{sec:discussion}

The example of probe D7-branes in the background of a stack of D3-branes illustrates that the spectrum of brane fluctuations can be modified quantitatively and qualitatively if the Hermiticity of the induced metric is not imposed and the determinants in the non-Abelian DBI action are not handled carefully. 

The corrected spectra we have found have some similarities with previous results \cite{Erdmenger:2007vj}, the main difference being that the degeneracy between scalar and vector modes is broken. We note that for relatively small values of $\lambda$, or for branes that are at very small separations, the mass of a heavy-light meson may lie in between the masses of heavy-heavy and light-light mesons. When this is the case, the picture of a heavy-light meson as a semiclassical string stretched between the two branes is probably breaking down. This is quite clear for two branes very close together, but perhaps less so for the case where the separation between the two branes is very large but $\lambda$ is not asymptotically large. This is perhaps indicating that the large 't Hooft limit and the limit of large quark mass separation are not commuting. Further study of the classical and quantum open string theory could shed more light on this issue.

From the point of view of QCD phenomenology, the D3-D7 model is not expected to provide a realistic description of the QCD meson spectra. Nevertheless, it does exhibit a range of non-Abelian symmetry-breaking phases~\cite{Erdmenger:2008yj,Ammon:2008fc,Ammon:2009fe,Erdmenger:2011hp,Chunlen:2012zy,Hoyos:2016ahj, Erdmenger:2023hkl}, and it would be interesting to revisit these constructions in light of our results. 
The existence of a very light vector meson at finite 't Hooft coupling in this model is worrisome for other holographic setups that aim at reproducing the confining phase of QCD more accurately. It would therefore be worthwhile to apply our technique to Sakai--Sugimoto model \cite{Sakai:2004cn} and to bottom-up models incorporating a DBI sector, such as V-QCD~\cite{Jarvinen:2011qe}.

Our analysis demonstrates that the method for extracting the meson spectrum from non-coincident probe D7-brane configurations is both robust and adaptable. While we have focused on the probe D7-brane setup in a supersymmetric D3-brane background, the essential features of our approach, namely linearizing the worldvolume fluctuations around a classical D7-brane profile and properly accounting for inter-brane separation, are general. 

In this work we restricted attention to fluctuations independent of the internal $S^3$ directions. 
A natural extension would be to include higher spherical harmonics on $S^3$, thereby completing the spectrum and allowing the study of angular excitations and their coupling patterns. %
Furthermore, as shown in the appendix \ref{app:pq}, it is straightforward to extend this framework to more general D$p$-D$q$-brane intersections~\cite{Arean:2006pk,Itsios:2016ffv}, including cases with reduced supersymmetry or more intricate internal geometry. Moreover, the setup is not restricted to conformally AdS backgrounds and could be applied to a broad class of holographic duals, including confining or deformed backgrounds. This could potentially allow systematic studies of meson spectra in more realistic QCD-like holographic models.

A natural direction for future work is to consider nonzero temperature and density effects by embedding the D7-brane probes in black hole geometries. This would allow for real-time spectral analysis of meson excitations at nonzero temperature or chemical potential, through the computation of retarded correlators and spectral functions~\cite{Erdmenger:2007ap,Erdmenger:2007ja}. These encode dynamical information such as quasiparticle widths, thermal broadening, and dissociation patterns, and eventually give access to weak decay rates~\cite{CruzRojas:2024etx,Hoyos:2024pkl} and transport coefficients~\cite{Hoyos:2020hmq,Hoyos:2021njg} of the flavor sector, especially in the neutron star context~\cite{Jarvinen:2021jbd,Hoyos:2021uff}. Depending on the precise embedding and background choice, one could study phenomena such as chiral symmetry restoration~\cite{Aharony:2006da}, meson melting thresholds~\cite{Hoyos-Badajoz:2006dzi}, or thermal screening lengths~\cite{Karch:2002xe}.

It would be especially interesting to explore the above questions in backgrounds with nontrivial dilaton flows, such as those appearing in the Liu--Tseytlin \cite{Liu:1999fc} and Constable--Myers geometry~\cite{Constable:1999ch}, which might capture aspects of renormalization group flow in the dual field theory. Our formalism can be adapted to these more general settings, offering a concrete route toward a more comprehensive holographic treatment of meson dynamics in strongly coupled, anisotropic~\cite{Azeyanagi:2009pr,Mateos:2011ix}, or thermal media.


\begin{acknowledgments}
We thank Francesco Bigazzi, Aldo L. Cotrone, Johanna Erdmenger, Nick Evans, Mauro Giliberti, and Matti J\"arvinen for suggestions, comments, and very helpful discussions. 
C.~H. has been supported in part by the Agencia Estatal de Investigaci\'on and the Ministerio de Ciencia, Innovaci\'on y Universidades through the Spanish grants PID2021-123021NB-I00 and PID2024-161500NB-I00. N.~J. has been supported in part by the Research Council of Finland grant no. 3545331. A.~O. wishes to thank the University of Helsinki and the King's College of London for kind hospitality during the preparation of this work. 
\end{acknowledgments}

\appendix

\section{Generalization of the DBI expansion to the D\texorpdfstring{$p$}{p}/D\texorpdfstring{$q$}{q} system}\label{app:pq}
In this appendix, we provide a generalization of the formulas we used in the main text to write the quadratic expansion of the D7-branes' DBI action to a generic D$p$/D$q$ system. Let us consider $\Nf$ D$q$-branes which probe the background generated by the backreaction of $\Nc$ D$p$-branes. The coordinates on the D$q$-branes worldvolumes are $\sigma^a$ with $a=0,\ldots,q$, while the embedding functions along the transverse directions are $X^I(\sigma)$ with $I=q+1,\ldots,9$. As in the main text, we choose the embeddings such that $X^a=\sigma\,\mathbb{1}_{\Nf}$. The non-Abelian gauge connection on the branes' worldvolume is $A_a(\sigma)$.

The D$q$-branes action involves the induced metric on the branes' worldvolume, which in its Hermition formulation reads
\begin{equation}
G_{ab}=P[g_{ab}]=G_{ab}(X)+\frac{1}{2}\left(D_a X^I G_{IJ}(X) D_b X^J+D_b X^I G_{IJ}(X) D_a X^J\right) \ ,
\end{equation}
where
\begin{equation}
    D_a X^I=\partial_a X^I+i[A_a,X^I]\ .
\end{equation}

A D$q$-brane also couples to the pullback of the background dilaton $\Phi=P[\phi]$ and to the Kalb--Ramond field $b_{MN}$ through the combination
\begin{equation}
E_{MN}=P[g_{MN}+b_{MN}] \ .
\end{equation}

If we have a non-zero Kalb--Ramond field, since the pullback is a linear operation on tensors, we have to require that the pullback on $b_{MN}$ is Hermitian, as for the metric tensor. This way, $E_{MN}$ is Hermitian. In the following, we will keep the discussion general and work directly with $E$. Moreover, if the pullback of the dilaton field is defined such that the result is Hermitian, then Hermiticity is respected throughout the whole action thanks to the cyclic property of the trace.

The Hermitian definition of the $Q$ matrix still reads
\begin{equation}
    Q^I\,_J = \delta^I_{\ J}\,\mathbb{1}_{\Nf}+\dfrac{1}{4\pi \ls^2}\big\{i[X^I,X^K],E_{KJ}\big\}\ .
\end{equation}

All in all, the Myers' action for $\Nf$ D$q$-branes reads \cite{Myers:1999ps}
\begin{equation}
S_{\text{D}q}=-T_q\int \dd^{q+1}\sigma\,\text{STr}_{\Nf} \left[e^{-\Phi}\sqrt{-\det\left(E_{ab}+E_{aI}(Q^{-1}-\delta \mathbb{1}_{\Nf})^{IJ}E_{Jb}+2\pi \ls^2 F_{ab}\right) \det Q^I_{\ J}} \right] \ ,
\end{equation}
where $T_q = 1/((2\pi)^q\gs\ls^{q+1})$, and $F_{ab} = \partial_aA_b-\partial_bA_a + i[A_a,A_b]$.

In general, the D$q$-branes can have a background embedding $\overline X\,^I$ and a non-zero background gauge field $\overline A_a$. These are diagonal matrices in the flavor indices. We now introduce fluctuations around the background solutions
\begin{align}
    &X^I= \overline{X}\,^I+\mathcal{X}^I \ ,& &A_a = \overline{A}_a + \mathcal{A}_a\ ,&
\end{align}
with components $x_{ij}$, and $(a_a)_{ij}$. 

The expansion of the induced matrix $E$ to quadratic order reads
\begin{equation}
    E_{ab}\sim E^{(0)}_{ab}+E^{(1)}_{ab}+E^{(2)}_{ab}\ .
\end{equation}
This involves the expansion of $E_{ab}(X)$ itself, and the expansion of the term involving the derivatives of the embedding functions. The $E$ matrix on the non-Abelian embedding can be computed through the following Taylor expansion
    \begin{align}\nonumber
    E_{MN}(X^{q+1},\dots,X^9) = &\sum\limits_{n_1=0}^\infty\cdots\sum_{n_{9-q}
    =0}^\infty\dfrac{1}{n_1!\cdots  n_{9-q}!}\dfrac{\partial^{n_1}}{\partial x_{q+1}^{n_1}}\cdots\dfrac{\partial^{n_{9-q}}}{\partial x_9^{n_{9-q}}}e_{MN}(x_{q+1},\dots,x_9)\bigg|_{x_{q+1}=\cdots =x_9=0} \\
    &\cdot \mathcal{S}[(X^{q+1})^{n_1}\cdots(X^{9})^{n_{9-q}}] \ ,
\end{align}
where $\mathcal{S}[(X^{q+1})^{n_1}\cdots(X^{9})^{n_{9-q}}]$ is the symmetrization operation.

As in the main text, this expansion can be truncated at second order in the fluctuations
\begin{equation}
    E_{ab}(X) \sim \overline E_{ab} + H^{(1)}_{ab}+H^{(2)}_{ab}\ ,
\end{equation}
where we have defined
\begin{align}
    E_{ab}(X^I=\overline{X}\,^I) = \overline E_{ab} \ .
\end{align}

Therefore, we can write the terms in the quadratic expansion of $E$ as
\begin{align}
    E^{(0)}_{ab} = \overline{E}_{ab}+\dfrac{1}{2}\left(\partial_a\overline X\,^I\,\overline{E}_{IJ}\partial_b\overline{X}\,^J+\partial_b\overline X\,^I\,\overline{E}_{IJ}\partial_a\overline{X}\,^J\right)
\end{align}
and
\begin{align}
    &E^{(1)}_{ab} = H_{ab}^{(1)}+\dfrac{1}{2}\bigg(\partial_a\overline X\,^I\,H_{IJ}^{(1)}\partial_b\overline{X}\,^J+\partial_b\overline X\,^I\,H_{IJ}^{(1)}\partial_a\overline{X}\,^J \\ \nonumber
    &+(\partial_a\mathcal{X}^I+i[\mathcal{A}_a,\overline{X}\,^I]+i[\overline{A}_a,\mathcal{X}^I])\overline{E}_{IJ}\partial_b\overline{X}\,^J+\partial_a\overline{X}\,^I\,\overline{E}_{IJ}(\partial_b\mathcal{X}^J+i[\mathcal{A}_b,\overline{X}\,^J]+i[\overline{A}_b,\mathcal{X}^J])\\ \nonumber
    &+(\partial_b\mathcal{X}^I+i[\mathcal{A}_b,\overline{X}\,^I]+i[\overline{A}_b,\mathcal{X}^I])\overline{E}_{IJ}\partial_a\overline{X}\,^J+\partial_b\overline{X}\,^I\,\overline{E}_{IJ}(\partial_a\mathcal{X}^J+i[\mathcal{A}_a,\overline{X}\,^J]+i[\overline{A}_a,\mathcal{X}^J])\bigg) \ ,
\end{align}
and
\begin{align}
    &E^{(2)}_{ab} = H_{ab}^{(2)}+\dfrac{1}{2}\bigg(\partial_a\overline X\,^I\,H_{IJ}^{(2)}\partial_b\overline{X}\,^J+\partial_b\overline X\,^I\,H_{IJ}^{(2)}\partial_a\overline{X}\,^J\\ \nonumber
     &+(\partial_a\mathcal{X}^I+i[\mathcal{A}_a,\overline{X}\,^I]+i[\overline{A}_a,\mathcal{X}^I])\overline{E}_{IJ}(\partial_b\mathcal{X}^J+i[\mathcal{A}_b,\overline{X}\,^J]+i[\overline{A}_b,\mathcal{X}^J])\\ \nonumber
     &+(\partial_b\mathcal{X}^I+i[\mathcal{A}_b,\overline{X}\,^I]+i[\overline{A}_b,\mathcal{X}^I])\overline{E}_{IJ}(\partial_a\mathcal{X}^J+i[\mathcal{A}_a,\overline{X}\,^J]+i[\overline{A}_a,\mathcal{X}^J])\\ \nonumber
     &+(\partial_a\mathcal{X}^I+i[\mathcal{A}_a,\overline{X}\,^I]+i[\overline{A}_a,\mathcal{X}^I])H^{(1)}_{IJ}\partial_b\overline{X}\,^J+\partial_a\overline{X}\,^I\,H^{(1)}_{IJ}(\partial_b\mathcal{X}^J+i[\mathcal{A}_b,\overline{X}\,^J]+i[\overline{A}_b,\mathcal{X}^J])\\ \nonumber
     &+(\partial_b\mathcal{X}^I+i[\mathcal{A}_b,\overline{X}\,^I]+i[\overline{A}_b,\mathcal{X}^I])H^{(1)}_{IJ}\partial_a\overline{X}\,^J+\partial_b\overline{X}\,^I\,H^{(1)}_{IJ}(\partial_a\mathcal{X}^J+i[\mathcal{A}_a,\overline{X}\,^J]+i[\overline{A}_a,\mathcal{X}^J])\\ \nonumber
     &+\partial_a\overline{X}\,^I\overline{E}_{IJ}\,i[\mathcal{A}_b,\mathcal{X}^J]+i[\mathcal{A}_a,\mathcal{X}^I]\overline{E}_{IJ}\partial_b\overline{X}\,^J + \partial_b\overline{X}\,^I\overline{E}_{IJ}\,i[\mathcal{A}_a,\mathcal{X}^J]+i[\mathcal{A}_b,\mathcal{X}^I]\overline{E}_{IJ}\partial_a\overline{X}\,^J\bigg) \ .
\end{align}

In a general background metric, a complication arises from the term $P[E_{aI}(Q^{-1}-\delta\,\mathbb{1}_{\Nf})^{IJ}E_{Jb}]$ that could contribute to the second-order expansion of the DBI action. We cannot rearrange this term without knowing the specifics of the system, and it has to be computed case by case. However, we can provide a general formula for computing the determinant $\det Q^I_{\ J}$
\begin{align}
   \det Q^I_{\ J}= \dfrac{1}{(9-q)!}\epsilon_{I_1I_2\cdots I_{9-q}}\epsilon^{J_1J_2\cdots J_{9-q}}Q^{I_1}_{\ J_1}Q^{I_2}_{\ J_2}\cdots Q^{I_{9-q}}_{\ J_{9-q}} \ .
\end{align}

Finally, we have to expand the determinant 
\begin{equation}
    \det(E_{ab}+E_{aI}(Q^{-1}-\delta\,\mathbb{1}_{\Nf})^{IJ}E_{Jb}+ 2\pi \ls^2 F_{ab}) \equiv \det K_{ab} = \dfrac{1}{(q+1)!}\epsilon^{a_0\cdots a_{q}}\epsilon^{b_0\cdots b_{q}}K_{a_0b_0}\cdots K_{a_qb_q}\,.
\end{equation}

As in the main text, the matrix $K_{ab}$ has the following expansion to quadratic order 
\begin{equation}
    K_{ab} \sim K^{(0)}_{ab}+\Xi^{(1)}_{ab}+\Xi^{(2)}_{ab},
\end{equation}
where now the terms of the expansion have a more complicated form, but are calculable with the formula we gave in this appendix. It is still true that non-zero components of the matrices in the above expansions are $(k^{(0)}_{ab})_{ii}$, $(\xi^{(1)}_{ab})_{ij}$, and $(\xi^{(2)}_{ab})_{ii}$. As a consequence, the determinant is expanded at quadratic order as 
\begin{equation}
    \det K_{ab}\sim \Delta_0+\Delta_1^{(1)}+\Delta_1^{(2)}+\Delta_2^{(2)}\ ,
\end{equation}
where
\begin{align}
&\Delta_0 = \det(K^{(0)}_{ab}) \\
&\Delta_1^{(n)} = \frac{1}{(q+1)!}\sum_{k=0}^q \epsilon^{a_0\cdots a_k\cdots a_q}\epsilon^{b_0\cdots b_k \cdots  b_q} K^{(0)}_{a_0 b_0}\cdots K^{(0)}_{a_{k-1} b_{k-1}} \Xi_{a_k b_k}^{(n)} K^{(0)}_{a_{k+1} b_{k+1}}\cdots K^{(0)}_{a_q b_q} \\ \nonumber
&\Delta_2^{(2)} = \frac{1}{(q+1)!}\sum_{k=1}^q\sum_{l=0}^{k-1} \epsilon^{a_0\cdots a_l\cdots  a_k\cdots a_q}\epsilon^{b_0\cdots b_l\cdots b_k \cdots  b_q} K^{(0)}_{a_0 b_0} \\
&\hspace{3cm}\,\,\,\,\cdots K^{(0)}_{a_{l-1} b_{l-1}} \Xi_{a_l b_l}^{(1)} K^{(0)}_{a_{l+1} b_{l+1}} \cdots K^{(0)}_{a_{k-1} b_{k-1}} \Xi_{a_k b_k}^{(1)} K^{(0)}_{a_{k+1} b_{k+1}}\cdots K^{(0)}_{a_q b_q}\ .
\end{align}

Finally, we can use the formula for the mixed determinant \eqref{eq:mixeddet}, which holds for general dimension, and write
\begin{align}
&(\Delta_1^{(1)})_{ij} =\frac{1}{(q+1)!}\sum_{k=0}^q \frac{\partial^{q+1}}{\partial t_0\cdots\partial t_q}\det\bigg[\sum_{n_i<k} t_{n_i}k^{(0)}_{ii}+t_k\xi^{(1)}_{ij}+\sum_{n_j>k} t_{n_j}k^{(0)}_{jj} \bigg] \\ 
&(\Delta_1^{(2)})_{ii} = \frac{1}{(q+1)!}\sum_{k=0}^q \frac{\partial^{q+1}}{\partial t_0\cdots\partial t_q}\det\bigg[\sum_{n_i<k} t_{n_i}k^{(0)}_{ii}+t_k\xi^{(2)}_{ii}+ \sum_{n_i'>k} t_{n_i'}k^{(0)}_{ii}\bigg] \\
&(\Delta_2^{(2)})_{ii} =\frac{1}{8!}\sum_{k=0}^{q-1}\sum_{l=k+1}^q \frac{\partial^{q+1}}{\partial t_0\cdots\partial t_q}\det\bigg[\sum_{n_i<k} t_{n_i}k^{(0)}_{ii}+t_k\xi^{(1)}_{ij} 
+\sum_{k<n_j<l} t_{n_j} k^{(0)}_{jj} +t_l \xi^{(1)}_{ji}+\sum_{n_i'>l} t_{n_i'}k^{(0)}_{ii}\bigg] \ .
\end{align}

Specializing these formulas to a specific D$p$/D$q$ system will provide a consistent expansion to the second order in the fluctuations of the non-Abelian flavor D$q$-branes' action.

We conclude this appendix by providing the Wess--Zumino action for the $\Nf$ D$q$-branes in a generic background. It reads \cite{Myers:1999ps}
\begin{equation}
    S_{\text{D}q}^\text{WZ} = T_q\int \text{STr}_{\Nf}\left(P\left[e^{i\,2\pi \ls^2\iota_{X}\iota_X}\left(\sum\limits_n C_ne^{b}\right)\right]e^{2\pi \ls^2F}\right)\ ,
\end{equation}
where we recall that $\iota_X$ denotes the interior product with the transverse matrix scalars. The Ramond--Ramond forms $C_n$ can be non-zero depending on the background generated by the backreaction of $\Nc$ D$p$-branes, while $b$ is the Kalb--Ramond field.

\section{Expansions for the numerical solutions}\label{app:numerics}

\subsection{Scalar fluctuation}\label{app:numericsscalar}

The first coefficients in the expansion around the origin \eqref{eq:scalarIRexpansion} are
\begin{align*}
    &a_2 = \frac{
  2 \left( -\Delta m^2 M_\text{meson} + \Delta m^3 + 4 \overline{m}^2 M_\text{meson} + 4 \overline{m}^2 \Delta m \right)
  \left( -4 \overline{m}^2 M_\text{meson} + 4 \overline{m}^2 \Delta m + \Delta m^2 M_\text{meson} + \Delta m^3 \right)
  \lambda}{
  \pi (-4 \overline{m}^2 + \Delta m^2)^2 (16 \overline{m}^4 + 24 \overline{m}^2 \Delta m^2 + \Delta m^4)}\,,
\end{align*}
\begin{align*}
  &a_4 =  \bigg( 4 \lambda \bigg[ -4\pi (4 \overline{m}^2 + \Delta m^2) \Big(
      512 \overline{m}^6 M_\text{meson}^2 \Delta m^2
      + 32 \overline{m}^2 M_\text{meson}^2 \Delta m^6
      - M_\text{meson}^2 \Delta m^8
      + \Delta m^{10} \\
    & \quad + 256 \overline{m}^8 (-M_\text{meson}^2 + \Delta m^2)
      + 224 \overline{m}^4 \Delta m^4 (-M_\text{meson}^2 + \Delta m^2)
    \Big) + \\
    &\quad\left( \Delta m^2 (-M_\text{meson} + \Delta m)
        + 4 \overline{m}^2 (M_\text{meson} + \Delta m) \right)^2\left( 4 \overline{m}^2 (-M_\text{meson} + \Delta m)
        + \Delta m^2 (M_\text{meson} + \Delta m) \right)^2 \lambda
  \bigg] \bigg) \\
  &\quad\times\bigg( 3 \pi^2 (-4 \overline{m}^2 + \Delta m^2)^4
    \left( 16 \overline{m}^4 + 24 \overline{m}^2 \Delta m^2 + \Delta m^4 \right)^2\bigg)^{-1} \ . \hspace{1.5cm}
\end{align*}
The first coefficients in the asymptotic expansion at the boundary \eqref{eq:scalarUVexpansion} are
\begin{align*}
    &b_4 = \dfrac{\lambda}{8\pi}(\Delta m^2-M_\text{meson}^2)\,, & \\
    &b_6 = \frac{ -128 \pi ^2 \Delta m^2 \overline m^2+4 \pi  \lambda  \left(\Delta m^2+4 \overline m^2\right) (M_\text{meson}^2-\Delta m^2)+\lambda ^2 \left(M_\text{meson}^2-\Delta m^2\right)^2}{192 \pi ^2}\ . & 
\end{align*}

\subsection{Transverse vector fluctuation}\label{app:numericsvector}

The first non-zero coefficients in the expansion around the origin given in \eqref{eq:vectorexpansions} are
\begin{align*}
    &c_2=\Big( 
    2 \big( 
        -430080 \overline m^{12}  M_\text{meson}^2 
        + 430080  \overline m^{12} \Delta m^2 
        - 501760  \overline m^{10} M_\text{meson}^2 \Delta m^2 
        - 645120  \overline m^{10} \Delta m^4 \\
&\quad 
        - 173824  \overline m^8 M_\text{meson}^2 \Delta m^4 
        + 403200  \overline m^8 \Delta m^6 
        - 99072  \overline m^6 M_\text{meson}^2 \Delta m^6 
        - 134400  \overline m^6 \Delta m^8 \\
&\quad 
        - 10864  \overline m^4 M_\text{meson}^2 \Delta m^8 
        + 25200  \overline m^4 \Delta m^{10} 
        - 1960  \overline m^2  M_\text{meson}^2 \Delta m^{10} 
        - 2520  \overline m^2 \Delta m^{12} \\
&\quad 
        - 105 M_\text{meson}^2 \Delta m^{12} 
        + 105 \Delta m^{14} 
    \big) \lambda \Big) \times \Big( 
    3 \pi (2 \overline m - \Delta m)^2 (2 \overline m + \Delta m)^2 \big( 
        143360  \overline m^{12} 
         \\
&\quad 
       + 71680  \overline m^{10} \Delta m^2 
        + 77056  \overline m^8 \Delta m^4 + 13568  \overline m^6 \Delta m^6 
        + 4816  \overline m^4 \Delta m^8 
        + 280  \overline m^2 \Delta m^{10} 
        + 35 \Delta m^{12} 
    \big) 
\Big)^{-1}\,,
\end{align*}
\begin{align*}
    &c_4= \Big( 
    4 \lambda \Big[ 
        2959500902400 \overline{m}^{26} M_\text{meson}^2 \pi 
        - 2959500902400 \overline{m}^{26} \pi\, \Delta m^2 
        + 4685876428800 \overline{m}^{24} M_\text{meson}^2 \pi\, \Delta m^2 \\
&\quad 
        + 11098128384000 \overline{m}^{24} \pi\, \Delta m^4 
        - 1504412958720 \overline{m}^{22} M_\text{meson}^2 \pi\, \Delta m^4 
        - 7176789688320 \overline{m}^{22} \pi\, \Delta m^6 \\
&\quad 
        - 657960468480 \overline{m}^{20} M_\text{meson}^2 \pi\, \Delta m^6 
        + 1757203660800 \overline{m}^{20} \pi\, \Delta m^8 
        - 1138989465600 \overline{m}^{18} M_\text{meson}^2 \pi\, \Delta m^8 \\
&\quad 
        - 728314675200 \overline{m}^{18} \pi\, \Delta m^{10} 
        - 442323173376 \overline{m}^{16} M_\text{meson}^2 \pi\, \Delta m^{10} 
        + 419069952000 \overline{m}^{16} \pi\, \Delta m^{12} \\
&\quad 
        - 219690565632 \overline{m}^{14} M_\text{meson}^2 \pi\, \Delta m^{12} 
        - 68785274880 \overline{m}^{14} \pi\, \Delta m^{14} 
        - 54922641408 \overline{m}^{12} M_\text{meson}^2 \pi\, \Delta m^{14} \\
&\quad 
        - 17196318720 \overline{m}^{12} \pi\, \Delta m^{16} 
        - 6911299584 \overline{m}^{10} M_\text{meson}^2 \pi\, \Delta m^{16} 
        + 6547968000 \overline{m}^{10} \pi\, \Delta m^{18} \\
&\quad 
        - 1112294400 \overline{m}^8 M_\text{meson}^2 \pi\, \Delta m^{18} 
        - 711244800 \overline{m}^8 \pi\, \Delta m^{20} 
        - 40158720 \overline{m}^6 M_\text{meson}^2 \pi\, \Delta m^{20} \\
&\quad 
        + 107251200 \overline{m}^6 \pi\, \Delta m^{22} 
        - 5738880 \overline{m}^4 M_\text{meson}^2 \pi\, \Delta m^{22} 
        - 27377280 \overline{m}^4 \pi\, \Delta m^{24} \\
&\quad 
        + 1117200 \overline{m}^2  M_\text{meson}^2 \pi\, \Delta m^{24} 
        + 2646000 \overline{m}^2 \pi\, \Delta m^{26} 
        + 44100 M_\text{meson}^2 \pi\, \Delta m^{26} 
        - 44100 \pi\, \Delta m^{28} \\
&\quad 
        + \Big( 
            430080 \overline{m}^{12} (M_\text{meson} - \Delta m)(M_\text{meson} + \Delta m) 
            + 105 (M_\text{meson} - \Delta m)\Delta m^{12}(M_\text{meson} + \Delta m) \\
&\quad 
            + 112 \overline{m}^4 \Delta m^8 (97 M_\text{meson}^2 - 225 \Delta m^2) 
            + 280 \overline{m}^2 \Delta m^{10}(7 M_\text{meson}^2 + 9 \Delta m^2) \\
&\quad 
            + 71680 \overline{m}^{10} (7 M_\text{meson}^2 \Delta m^2 + 9 \Delta m^4) 
            + 1792 \overline{m}^8 (97 M_\text{meson}^2 \Delta m^4 - 225 \Delta m^6) \\
&\quad 
            + 768 \overline{m}^6 (129 M_\text{meson}^2 \Delta m^6 + 175 \Delta m^8)
        \Big)^2 
    \Big]
\Big)\Big( 
    27 \pi^2 (-4 \overline{m}^2 + \Delta m^2)^4 
    \Big( 
        143360 \overline{m}^{12} 
        + 71680 \overline{m}^{10} \Delta m^2 
        \\
&\quad  + 77056 \overline{m}^8 \Delta m^4
        + 13568 \overline{m}^6 \Delta m^6 
        + 4816 \overline{m}^4 \Delta m^8 
        + 280 \overline{m}^2 \Delta m^{10} 
        + 35 \Delta m^{12}
    \Big)^2 
\Big)^{-1} \ .\qquad\quad
\end{align*}
The first nonzero coefficients in the asymptotic boundary expansion given in \eqref{eq:vectorexpansions} are
\begin{align*}
   &d_4 = \dfrac{\lambda}{8\pi}(\Delta m^2-M_\text{meson}^2) \\
   &d_6 =  \frac{ -128 \pi ^2 \Delta m^2 \overline m^2+4 \pi  \lambda  \left(\Delta m^2+4 \overline m^2\right) (M_\text{meson}^2-\Delta m^2)+\lambda ^2 \left(M_\text{meson}^2-\Delta m^2\right)^2}{192 \pi ^2} \ . & 
\end{align*}

\section{Numerical analysis of the equation of motion for transverse vector fluctuations}\label{app:numanalysis}
In this appendix, we want to shed light on the dependence of the meson spectrum on the values of $m_\text{heavy}$ when it takes large values. We have to distinguish two limits that accomplish this scenario, which are $m_\text{heavy}\to\infty$ (IR limit) and $m_\text{light} \to 0$ (UV limit).\footnote{We remember that we have defined $m_i \equiv m_\text{heavy}$, and $m_j \equiv m_\text{light}$.}

Let us consider first the $m_\text{heavy}\to\infty$ limit of the equation \eqref{eq:eqphiij}. We get
\begin{equation}
    \partial_v^2\phi^\text{IR}+\dfrac{3}{v}\dfrac{1-v^2}{1+v^2}\partial_v\phi^\text{IR}+\dfrac{\widetilde M_\text{meson}^2}{1+v^2}\phi^\text{IR} = 0 \ ,
\end{equation}
where we performed the redefinitions
\begin{align}
    &\rho\to m_\text{light}\,v,& &M_\text{meson}\to \widetilde M_\text{meson}\,m_\text{heavy}\,\sqrt{\dfrac{3\pi}{4\lambda}}\ , &
\end{align}
and we have labeled the function which solves the equation of motion in the IR limit as $\phi^\text{IR}$.

The above equation can be solved analytically. If we select the regular solution at the origin of the spacetime $v=0$, we end up with
\begin{equation}
    \phi^\text{IR} =  \,_2F_1\left(-1-\sqrt{1-\frac{\widetilde M_\text{meson}^2}{4}},-1+\sqrt{1-\frac{\widetilde M_\text{meson}^2}{4}},2;-v^2\right) \ , 
\end{equation}
which goes to 1 for $v\to 0$. Let us observe that in the massless case $\widetilde M_\text{meson} = 0$, the solution trivializes to $\phi^\text{IR} =1$. 

The expansion of the solution close to the boundary $v\to\infty$ reads
\begin{align}\label{eq:nudef}
    &\phi^\text{IR}\sim a_-\,v^{2-2\nu}+a_+\,v^{2+2\nu}\,,& &\nu =\sqrt{1-\frac{\widetilde M_\text{meson}^2}{4 }}\ , &
\end{align}
where $a_-$ and $a_+$ are
\begin{align}\label{eq:apmcoeff}
    &a_- = \frac{\Gamma (-2 \nu )}{\Gamma (-1 -\nu) \Gamma (3-\nu )}\,,&  &a_+ = \frac{\,\Gamma (2 \nu )}{\Gamma (-1+\nu) \Gamma (3+\nu )} \  . &
\end{align}

These are the coefficients we have to match with the expansion close to the spacetime infrared region of the solution to the equation of motion in the $m_\text{light}\to 0$ limit.\footnote{Note that in the massless limit, $\widetilde M_\text{meson}=0$, the coefficients reduce to $a_-= 1/2$ and $a_+=0$. The two modes combine to give the constant solution.} The equation in this limit reads
\begin{align}
    &\partial_u^2\phi^\text{UV}+\left(\frac{-14 u^4-23 u^2+70 \left(u^2+1\right) \left(6 u^6+18 u^4+17 u^2+6\right) u^6-3}{u \left(u^2+1\right) \left(14 \left(u^2+1\right) \left(5 \left(2 u^8+4 u^6+3 u^4+u^2\right)+1\right) u^2+1\right)}\right)\partial_u\phi^\text{UV}\\ \nonumber
    &+ \left(\frac{\pi\,\widetilde M_\text{meson}^2 \left(7 \left(u^2+1\right) \left(5 \left(u^2+1\right) \left(3 \left(u^4+u^2\right)+2\right) u^2+2\right) u^2+1\right)-140\, \lambda \, u^6 \left(u^2+1\right)^3}{\pi\,  u^2 \left(u^2+1\right) \left(14 \left(u^2+1\right) \left(5 \left(2 u^8+4 u^6+3 u^4+u^2\right)+1\right) u^2+1\right)}\right)\phi^\text{UV} =0 \ ,
\end{align}
where we have performed the following redefinitions
\begin{align}
    &\rho\to m_\text{heavy}\,u,& &M_\text{meson}\to \widetilde M_\text{meson}\,m_\text{heavy}\,\sqrt{\dfrac{3\pi}{4\lambda}}\ , &
\end{align}
and we named $\phi^\text{UV}$ the solution.

Unfortunately, there is no analytic solution to this equation, and therefore, we have to resort to numerics. As in the main text, we employ the shooting method to solve the equation of motion in the bulk, using as cutoffs $u=10^{-5}$ and $u=100$. We must select the normalizable solution at the boundary in order to describe a mesonic excitation, and therefore, we shoot the solution from the boundary by employing the following expansion at the boundary
\begin{align}
    \phi^\text{UV}_\text{norm.}\bigg|_{u\to\infty} = \sum_{n=1}^{N_u}p_{2n}u^{-2n} \ ,
\end{align}
where $N_u = 6$, and by setting $p_2=1$, the first coefficients are
\begin{align*}
    &p_4 = \frac{4 \lambda -3 \pi\, \widetilde M_\text{meson}^2}{32 \pi }& &p_6 =\frac{16 \lambda ^2-8 \pi  \lambda  \left(3 \widetilde M_\text{meson}^2+16\right)+\pi ^2 \left(9 \widetilde M_\text{meson}^4+96 \widetilde M_\text{meson}^2-512\right)}{3072 \pi ^2}\ ,& &\ldots&
\end{align*}

Now, we have to shoot from the origin $u\to 0$. The general solution is a combination of two modes, namely $\varphi_\pm^\text{UV}(u)$, which match the expansion of the $m_\text{heavy}\to \infty$ solution for $u\to 0$
\begin{align}
    \varphi_\pm^\text{UV}(u)\sim u^{2\pm2\nu}\left(c_\pm + \dots\right) \ ,
\end{align}
where we $\nu$ is defined in \eqref{eq:nudef} as $\nu = \sqrt{1-\widetilde M^2_\text{meson}/4}$. We can write one equation of motion for each mode once we have extracted the near-origin behavior of the two modes
\begin{equation}
    \varphi_\pm^\text{UV}(u) = u^{2\pm2\nu}\varpi_\pm^\text{UV}(u) \ .
\end{equation}
Then, the equations for $\varpi_\pm(u)$ read
\begin{align}\nonumber
    &\partial_u^2\varpi_\pm^\text{UV} + \left(\dfrac{1}{u}\pm\frac{6 u}{u^2+1}\frac{2 \sqrt{4-\widetilde M^2_\text{meson}}}{u}+\frac{28 \left(2 u^2+1\right) \left(10 \left(3 u^8+6 u^6+4 u^4+u^2\right)+1\right) u}{14 \left(u^2+1\right) \left(5 \left(2 u^8+4 u^6+3 u^4+u^2\right)+1\right) u^2+1}\right)\partial_u\varpi_\pm^\text{UV} \\ \nonumber
    &+\bigg[-\pi \, \widetilde M_\text{meson}^2 \left(7 \left(u^2+1\right) \left(5 \left(u^2+1\right) \left(4 u^4+5 u^2+3\right) u^2+2\right) u^2+1\right) -140 \lambda  \left(u^2+1\right)^3 u^4 \\ \nonumber
    &+2 \pi  \left(\pm\sqrt{4-\widetilde M_\text{meson}^2}+2\right) \left(14 \left(54 u^2+5 \left(6 u^6+24 u^4+37 u^2+28\right) u^4+10\right) u^2+11\right)\bigg] \\
    &\cdot\big[\pi  \left(u^2+1\right) \left(14 \left(u^2+1\right) \left(5 \left(2 u^8+4 u^6+3 u^4+u^2\right)+1\right) u^2+1\right)\big]^{-1}\,\varpi_\pm^\text{UV} = 0 \ .
\end{align}

For the functions $\varpi_\pm^\text{UV}(u)$ we use the following expansion close to the boundary
\begin{equation}
    \varpi_\pm^\text{UV}\bigg|_{u\to 0} = \sum\limits_{n=0}^{N_\pm}e_{\pm,2n}\,u^{2n},
\end{equation}
where $N_\pm = 5$, and by setting $e_{\pm,0}=1$ the coefficients read
\begin{align*}
    &e_{\pm,2} = \dfrac{\widetilde M^2_\text{meson}-22\left(2\pm\sqrt{4-\widetilde M^2_\text{meson}}\right)}{4\left(1\pm\sqrt{4-\widetilde M^2_\text{meson}}\right)}\,,&\\
    &e_{\pm,4}= \frac{\widetilde M_\text{meson}^4-\left(682\pm45 \sqrt{4-\widetilde M_\text{meson}^2}\right) \widetilde M_\text{meson}^2\pm1332 \sqrt{4-\widetilde M_\text{meson}^2}+840}{32 \left(3-\widetilde M_\text{meson}^2\right)}\ ,& &\ldots&
\end{align*}
Note that in the $\widetilde M_\text{meson}\to 0$ limit, the coefficients become
\begin{align}
    &e_{+,2} =-\dfrac{22}{3},& &e_{-,2}=0,& &e_{+,4} =\dfrac{73}{4},& &e_{-,4} =-19\ ,& &\ldots&
\end{align}
Then, we can write the full solution and do the matching
\begin{align}
    \phi^\text{UV}(u) = c_\text{norm.}\,\phi_\text{norm.}^\text{UV}(u) = b_+\,u^{2+2\nu}\varpi_+^\text{UV}(u)+b_-\,u^{2-2\nu}\varpi_-^\text{UV}(u),
\end{align}
where the $b_\pm$ coefficients can be matched with the $\rho\to\infty$ (or $v\to\infty$) UV expansion of the IR solution $\phi^\text{IR}$ as follows
\begin{align}\label{eq:bpmcoeff}
    b_\pm = \left(\dfrac{m_\text{heavy}}{m_\text{light}}\right)^{2\pm 2\nu}a_\pm \ ,
\end{align}
where $a_\pm$ are defined in \eqref{eq:apmcoeff}. We can compute the Wronskian of the solutions
\begin{align}\nonumber
    \mathcal{W}&[\lambda,m_\text{light},m_\text{heavy},\widetilde M_\text{meson};u] =-\phi^\text{UV}_\text{norm.}(u)\Bigg(a_+\,\left(\dfrac{m_\text{heavy}}{m_\text{light}}\right)^{2+2\nu}\partial_u\left(u^{2+2\nu}\,\varpi_+^\text{UV}(u)\right)\\ \nonumber
    &+a_-\,\left(\dfrac{m_\text{heavy}}{m_\text{light}}\right)^{2-2\nu}\partial_u\left(u^{2-2\nu}\,\varpi_-^\text{UV}(u)\right)\Bigg)\\
    &+ \partial_u\phi^\text{UV}_\text{norm.}(u)\left(a_+\,\left(\dfrac{m_\text{heavy}}{m_\text{light}}\right)^{2+2\nu}u^{2+2\nu}\,\varpi_+^\text{UV}(u)+a_-\,\left(\dfrac{m_\text{heavy}}{m_\text{light}}\right)^{2-2\nu}u^{2-2\nu}\,\varpi_-^\text{UV}(u)\right) \ ,
\end{align}
and we use the \textbf{FindRoot} routine of \textit{Mathematica} to find the zeros of the Wronskian $\mathcal{W}$ at fixed $m_\text{light}$, and $m_\text{heavy}$. These zeros happen to be located at the values of $\widetilde M_\text{meson}$ for which the shot solutions smoothly match at an arbitrary value of the holographic coordinate $u=u_\text{match}$. For the numerics, we used $u_\text{match}=15$ and we checked that the results are independent of this value.

We varied the ratio of the heavy and light quark masses to find the values of the meson mass for which the Wronskian is zero. The results are displayed in figure \ref{fig:largeheavyquarkmass} where the meson mass (normalized by $\lambda/(\mathbf{m}^2_\mathbf{{q,\text{light}}}\pi)$) has been numerically computed for the following values of the ratio $\mathbf{m_{q,\text{heavy}}}/\mathbf{m_{q,\text{light}}}=\{50,75,100,150,200,250,300,400,500\}$ (red dots). The blue solid line corresponds to a fit with a power law function to extract the precise dependence of the meson mass on the heavy quark mass. In particular, if we restore the correct units, we have
\begin{align}
    \dfrac{\lambda}{\pi}\,\mathbf{M}^2_\text{meson} \approx a^2\,\dfrac{\mathbf{m}^4_\mathbf{{q,\text{light}}}}{\mathbf{m}^2_\mathbf{{q,\text{heavy}}}},
\end{align}
where $a\approx 288$, indicating that the meson mass decreases with $\mathbf{m_{q,\text{heavy}}}$.\footnote{We recall that $m_\text{heavy} = \sqrt{2\pi} \,\ls\,\mathbf{m_{q,\text{heavy}}}$.}

As a check for this behavior, we look at the coefficients $b_\pm$, see equation \eqref{eq:bpmcoeff}, when the meson mass goes like $m_\text{heavy}^{-1}$
\begin{align}
    &b_- = \dfrac{1}{2} + \mathcal{O}\left(\dfrac{1}{m_\text{heavy}}\right),& &b_+ = -c +\mathcal{O}\left(\dfrac{1}{m_\text{heavy}}\right),&
\end{align}
where $c\approx 10.66$ is found using the value $a\approx 288$ of the meson mass fit. This means that even in the limit of $m_\text{heavy}\to \infty$ where the fluctuation mode becomes massless, the coefficient $b_+$ does not go to zero but remains constant, allowing for a massless normalizable solution to the equation of motion.

\bibliographystyle{JHEP}
\bibliography{Ref}

\providecommand{\href}[2]{#2}\begingroup\raggedright\begin{thebibliography}{10}

\bibitem{Karch:2002sh}
A.~Karch and E.~Katz, \emph{{Adding flavor to AdS / CFT}},
  \href{https://doi.org/10.1088/1126-6708/2002/06/043}{\emph{JHEP} {\bfseries
  06} (2002) 043} [\href{https://arxiv.org/abs/hep-th/0205236}{{\ttfamily
  hep-th/0205236}}].

\bibitem{Kruczenski:2003be}
M.~Kruczenski, D.~Mateos, R.~C. Myers and D.~J. Winters, \emph{{Meson
  spectroscopy in AdS / CFT with flavor}},
  \href{https://doi.org/10.1088/1126-6708/2003/07/049}{\emph{JHEP} {\bfseries
  07} (2003) 049} [\href{https://arxiv.org/abs/hep-th/0304032}{{\ttfamily
  hep-th/0304032}}].

\bibitem{Kobayashi:2006sb}
S.~Kobayashi, D.~Mateos, S.~Matsuura, R.~C. Myers and R.~M. Thomson,
  \emph{{Holographic phase transitions at finite baryon density}},
  \href{https://doi.org/10.1088/1126-6708/2007/02/016}{\emph{JHEP} {\bfseries
  02} (2007) 016} [\href{https://arxiv.org/abs/hep-th/0611099}{{\ttfamily
  hep-th/0611099}}].

\bibitem{Sakai:2004cn}
T.~Sakai and S.~Sugimoto, \emph{{Low energy hadron physics in holographic
  QCD}}, \href{https://doi.org/10.1143/PTP.113.843}{\emph{Prog. Theor. Phys.}
  {\bfseries 113} (2005) 843}
  [\href{https://arxiv.org/abs/hep-th/0412141}{{\ttfamily hep-th/0412141}}].

\bibitem{Jarvinen:2011qe}
M.~J\"arvinen and E.~Kiritsis, \emph{{Holographic Models for QCD in the
  Veneziano Limit}}, \href{https://doi.org/10.1007/JHEP03(2012)002}{\emph{JHEP}
  {\bfseries 03} (2012) 002} [\href{https://arxiv.org/abs/1112.1261}{{\ttfamily
  1112.1261}}].

\bibitem{Leigh:1989jq}
R.~G. Leigh, \emph{{Dirac-Born-Infeld Action from Dirichlet Sigma Model}},
  \href{https://doi.org/10.1142/S0217732389003099}{\emph{Mod. Phys. Lett. A}
  {\bfseries 4} (1989) 2767}.

\bibitem{Myers:1999ps}
R.~C. Myers, \emph{{Dielectric branes}},
  \href{https://doi.org/10.1088/1126-6708/1999/12/022}{\emph{JHEP} {\bfseries
  12} (1999) 022} [\href{https://arxiv.org/abs/hep-th/9910053}{{\ttfamily
  hep-th/9910053}}].

\bibitem{Erdmenger:2007vj}
J.~Erdmenger, K.~Ghoroku and I.~Kirsch, \emph{{Holographic heavy-light mesons
  from non-Abelian DBI}},
  \href{https://doi.org/10.1088/1126-6708/2007/09/111}{\emph{JHEP} {\bfseries
  09} (2007) 111} [\href{https://arxiv.org/abs/0706.3978}{{\ttfamily
  0706.3978}}].

\bibitem{Erdmenger:2006bg}
J.~Erdmenger, N.~Evans and J.~Grosse, \emph{{Heavy-light mesons from the
  AdS/CFT correspondence}},
  \href{https://doi.org/10.1088/1126-6708/2007/01/098}{\emph{JHEP} {\bfseries
  01} (2007) 098} [\href{https://arxiv.org/abs/hep-th/0605241}{{\ttfamily
  hep-th/0605241}}].

\bibitem{Tseytlin:1997csa}
A.~A. Tseytlin, \emph{{On nonAbelian generalization of Born-Infeld action in
  string theory}},
  \href{https://doi.org/10.1016/S0550-3213(97)00354-4}{\emph{Nucl. Phys. B}
  {\bfseries 501} (1997) 41}
  [\href{https://arxiv.org/abs/hep-th/9701125}{{\ttfamily hep-th/9701125}}].

\bibitem{Hashimoto:1997gm}
A.~Hashimoto and W.~Taylor, \emph{{Fluctuation spectra of tilted and
  intersecting D-branes from the Born-Infeld action}},
  \href{https://doi.org/10.1016/S0550-3213(97)00399-4}{\emph{Nucl. Phys. B}
  {\bfseries 503} (1997) 193}
  [\href{https://arxiv.org/abs/hep-th/9703217}{{\ttfamily hep-th/9703217}}].

\bibitem{Koerber:2002zb}
P.~Koerber and A.~Sevrin, \emph{{The NonAbelian D-brane effective action
  through order alpha-prime**4}},
  \href{https://doi.org/10.1088/1126-6708/2002/10/046}{\emph{JHEP} {\bfseries
  10} (2002) 046} [\href{https://arxiv.org/abs/hep-th/0208044}{{\ttfamily
  hep-th/0208044}}].

\bibitem{Keurentjes:2004tu}
A.~Keurentjes, P.~Koerber, S.~Nevens, A.~Sevrin and A.~Wijns, \emph{{Towards an
  effective action for D-branes}},
  \href{https://doi.org/10.1002/prop.200410225}{\emph{Fortsch. Phys.}
  {\bfseries 53} (2005) 599}
  [\href{https://arxiv.org/abs/hep-th/0412271}{{\ttfamily hep-th/0412271}}].

\bibitem{2013arXiv1306.1315A}
S.~{Artstein-Avidan}, D.~{Florentin} and Y.~{Ostrover}, \emph{{Remarks about
  Mixed Discriminants and Volumes}},
  \href{https://doi.org/10.1142/S0219199713500314}{\emph{Commun. Contemp.
  Math.} {\bfseries 16} (2014) 1350031}
  [\href{https://arxiv.org/abs/1306.1315}{{\ttfamily 1306.1315}}].

\bibitem{2018arXiv180605105B}
A.~{Barvinok}, \emph{{Stability and complexity of mixed discriminants}},
  \href{https://doi.org/10.48550/arXiv.1806.05105}{\emph{Math. Comp. 89 (2020),
  717-735} (2018) arXiv:1806.05105}
  [\href{https://arxiv.org/abs/1806.05105}{{\ttfamily 1806.05105}}].

\bibitem{Erdmenger:2008yj}
J.~Erdmenger, M.~Kaminski, P.~Kerner and F.~Rust, \emph{{Finite baryon and
  isospin chemical potential in AdS/CFT with flavor}},
  \href{https://doi.org/10.1088/1126-6708/2008/11/031}{\emph{JHEP} {\bfseries
  11} (2008) 031} [\href{https://arxiv.org/abs/0807.2663}{{\ttfamily
  0807.2663}}].

\bibitem{Ammon:2008fc}
M.~Ammon, J.~Erdmenger, M.~Kaminski and P.~Kerner, \emph{{Superconductivity
  from gauge/gravity duality with flavor}},
  \href{https://doi.org/10.1016/j.physletb.2009.09.029}{\emph{Phys. Lett. B}
  {\bfseries 680} (2009) 516}
  [\href{https://arxiv.org/abs/0810.2316}{{\ttfamily 0810.2316}}].

\bibitem{Ammon:2009fe}
M.~Ammon, J.~Erdmenger, M.~Kaminski and P.~Kerner, \emph{{Flavor
  Superconductivity from Gauge/Gravity Duality}},
  \href{https://doi.org/10.1088/1126-6708/2009/10/067}{\emph{JHEP} {\bfseries
  10} (2009) 067} [\href{https://arxiv.org/abs/0903.1864}{{\ttfamily
  0903.1864}}].

\bibitem{Erdmenger:2011hp}
J.~Erdmenger, V.~Grass, P.~Kerner and T.~H. Ngo, \emph{{Holographic
  Superfluidity in Imbalanced Mixtures}},
  \href{https://doi.org/10.1007/JHEP08(2011)037}{\emph{JHEP} {\bfseries 08}
  (2011) 037} [\href{https://arxiv.org/abs/1103.4145}{{\ttfamily 1103.4145}}].

\bibitem{Chunlen:2012zy}
S.~Chunlen, K.~Peeters, P.~Vanichchapongjaroen and M.~Zamaklar,
  \emph{{Instability of N=2 gauge theory in compact space with an isospin
  chemical potential}},
  \href{https://doi.org/10.1007/JHEP01(2013)035}{\emph{JHEP} {\bfseries 01}
  (2013) 035} [\href{https://arxiv.org/abs/1210.6188}{{\ttfamily 1210.6188}}].

\bibitem{Hoyos:2016ahj}
C.~Hoyos, G.~Itsios and O.~Vasilakis, \emph{{Baryon Superfluids in AdS/CFT with
  Flavor}}, \href{https://doi.org/10.1007/JHEP01(2017)139}{\emph{JHEP}
  {\bfseries 01} (2017) 139}
  [\href{https://arxiv.org/abs/1611.07029}{{\ttfamily 1611.07029}}].

\bibitem{Erdmenger:2023hkl}
J.~Erdmenger, N.~Evans, Y.~Liu and W.~Porod, \emph{{Holographic Non-Abelian
  Flavour Symmetry Breaking}},
  \href{https://doi.org/10.3390/universe9060289}{\emph{Universe} {\bfseries 9}
  (2023) 289} [\href{https://arxiv.org/abs/2304.09190}{{\ttfamily
  2304.09190}}].

\bibitem{Arean:2006pk}
D.~Arean and A.~V. Ramallo, \emph{{Open string modes at brane intersections}},
  \href{https://doi.org/10.1088/1126-6708/2006/04/037}{\emph{JHEP} {\bfseries
  04} (2006) 037} [\href{https://arxiv.org/abs/hep-th/0602174}{{\ttfamily
  hep-th/0602174}}].

\bibitem{Itsios:2016ffv}
G.~Itsios, N.~Jokela and A.~V. Ramallo, \emph{{Collective excitations of
  massive flavor branes}},
  \href{https://doi.org/10.1016/j.nuclphysb.2016.06.008}{\emph{Nucl. Phys. B}
  {\bfseries 909} (2016) 677}
  [\href{https://arxiv.org/abs/1602.06106}{{\ttfamily 1602.06106}}].

\bibitem{Erdmenger:2007ap}
J.~Erdmenger, M.~Kaminski and F.~Rust, \emph{{Isospin diffusion in thermal
  AdS/CFT with flavor}},
  \href{https://doi.org/10.1103/PhysRevD.76.046001}{\emph{Phys. Rev. D}
  {\bfseries 76} (2007) 046001}
  [\href{https://arxiv.org/abs/0704.1290}{{\ttfamily 0704.1290}}].

\bibitem{Erdmenger:2007ja}
J.~Erdmenger, M.~Kaminski and F.~Rust, \emph{{Holographic vector mesons from
  spectral functions at finite baryon or isospin density}},
  \href{https://doi.org/10.1103/PhysRevD.77.046005}{\emph{Phys. Rev. D}
  {\bfseries 77} (2008) 046005}
  [\href{https://arxiv.org/abs/0710.0334}{{\ttfamily 0710.0334}}].

\bibitem{CruzRojas:2024etx}
J.~Cruz~Rojas, T.~Gorda, C.~Hoyos, N.~Jokela, M.~J{\"a}rvinen, A.~Kurkela
  et~al., \emph{{Estimate for the Bulk Viscosity of Strongly Coupled Quark
  Matter Using Perturbative QCD and Holography}},
  \href{https://doi.org/10.1103/PhysRevLett.133.071901}{\emph{Phys. Rev. Lett.}
  {\bfseries 133} (2024) 071901}
  [\href{https://arxiv.org/abs/2402.00621}{{\ttfamily 2402.00621}}].

\bibitem{Hoyos:2024pkl}
C.~Hoyos, A.~Olzi and D.~Rodriguez-Fernandez, \emph{{Weak rates in strongly
  coupled cold quark matter}},
  \href{https://doi.org/10.1007/JHEP12(2024)058}{\emph{JHEP} {\bfseries 12}
  (2024) 058} [\href{https://arxiv.org/abs/2407.21643}{{\ttfamily
  2407.21643}}].

\bibitem{Hoyos:2020hmq}
C.~Hoyos, N.~Jokela, M.~J\"arvinen, J.~G. Subils, J.~Tarrio and A.~Vuorinen,
  \emph{{Transport in strongly coupled quark matter}},
  \href{https://doi.org/10.1103/PhysRevLett.125.241601}{\emph{Phys. Rev. Lett.}
  {\bfseries 125} (2020) 241601}
  [\href{https://arxiv.org/abs/2005.14205}{{\ttfamily 2005.14205}}].

\bibitem{Hoyos:2021njg}
C.~Hoyos, N.~Jokela, M.~J{\"a}rvinen, J.~G. Subils, J.~Tarrio and A.~Vuorinen,
  \emph{{Holographic approach to transport in dense QCD matter}},
  \href{https://doi.org/10.1103/PhysRevD.105.066014}{\emph{Phys. Rev. D}
  {\bfseries 105} (2022) 066014}
  [\href{https://arxiv.org/abs/2109.12122}{{\ttfamily 2109.12122}}].

\bibitem{Jarvinen:2021jbd}
M.~J{\"a}rvinen, \emph{{Holographic modeling of nuclear matter and neutron
  stars}}, \href{https://doi.org/10.1140/epjc/s10052-022-10227-x}{\emph{Eur.
  Phys. J. C} {\bfseries 82} (2022) 282}
  [\href{https://arxiv.org/abs/2110.08281}{{\ttfamily 2110.08281}}].

\bibitem{Hoyos:2021uff}
C.~Hoyos, N.~Jokela and A.~Vuorinen, \emph{{Holographic approach to compact
  stars and their binary mergers}},
  \href{https://doi.org/10.1016/j.ppnp.2022.103972}{\emph{Prog. Part. Nucl.
  Phys.} {\bfseries 126} (2022) 103972}
  [\href{https://arxiv.org/abs/2112.08422}{{\ttfamily 2112.08422}}].

\bibitem{Aharony:2006da}
O.~Aharony, J.~Sonnenschein and S.~Yankielowicz, \emph{{A Holographic model of
  deconfinement and chiral symmetry restoration}},
  \href{https://doi.org/10.1016/j.aop.2006.11.002}{\emph{Annals Phys.}
  {\bfseries 322} (2007) 1420}
  [\href{https://arxiv.org/abs/hep-th/0604161}{{\ttfamily hep-th/0604161}}].

\bibitem{Hoyos-Badajoz:2006dzi}
C.~Hoyos-Badajoz, K.~Landsteiner and S.~Montero, \emph{{Holographic meson
  melting}}, \href{https://doi.org/10.1088/1126-6708/2007/04/031}{\emph{JHEP}
  {\bfseries 04} (2007) 031}
  [\href{https://arxiv.org/abs/hep-th/0612169}{{\ttfamily hep-th/0612169}}].

\bibitem{Karch:2002xe}
A.~Karch, E.~Katz and N.~Weiner, \emph{{Hadron masses and screening from AdS
  Wilson loops}},
  \href{https://doi.org/10.1103/PhysRevLett.90.091601}{\emph{Phys. Rev. Lett.}
  {\bfseries 90} (2003) 091601}
  [\href{https://arxiv.org/abs/hep-th/0211107}{{\ttfamily hep-th/0211107}}].

\bibitem{Liu:1999fc}
H.~Liu and A.~A. Tseytlin, \emph{{D3-brane D instanton configuration and N=4
  superYM theory in constant selfdual background}},
  \href{https://doi.org/10.1016/S0550-3213(99)00259-X}{\emph{Nucl. Phys. B}
  {\bfseries 553} (1999) 231}
  [\href{https://arxiv.org/abs/hep-th/9903091}{{\ttfamily hep-th/9903091}}].

\bibitem{Constable:1999ch}
N.~R. Constable and R.~C. Myers, \emph{{Exotic scalar states in the AdS / CFT
  correspondence}},
  \href{https://doi.org/10.1088/1126-6708/1999/11/020}{\emph{JHEP} {\bfseries
  11} (1999) 020} [\href{https://arxiv.org/abs/hep-th/9905081}{{\ttfamily
  hep-th/9905081}}].

\bibitem{Azeyanagi:2009pr}
T.~Azeyanagi, W.~Li and T.~Takayanagi, \emph{{On String Theory Duals of
  Lifshitz-like Fixed Points}},
  \href{https://doi.org/10.1088/1126-6708/2009/06/084}{\emph{JHEP} {\bfseries
  06} (2009) 084} [\href{https://arxiv.org/abs/0905.0688}{{\ttfamily
  0905.0688}}].

\bibitem{Mateos:2011ix}
D.~Mateos and D.~Trancanelli, \emph{{The anisotropic N=4 super Yang-Mills
  plasma and its instabilities}},
  \href{https://doi.org/10.1103/PhysRevLett.107.101601}{\emph{Phys. Rev. Lett.}
  {\bfseries 107} (2011) 101601}
  [\href{https://arxiv.org/abs/1105.3472}{{\ttfamily 1105.3472}}].

\end{thebibliography}\endgroup

\end{document}